\pdfoutput=1

\documentclass[3p]{elsarticle}
\usepackage[unicode,pagebackref=true]{hyperref}
\renewcommand*\backref[1]{\ifx#1\relax \else (#1) \fi}
\usepackage[dvipsnames]{xcolor}
\usepackage{graphicx}
\usepackage{amssymb}
\RequirePackage{mathptmx}
\RequirePackage[T1]{fontenc}
\usepackage{heppennames2}
\usepackage{lhchiggs}
\usepackage{cernunits}
\usepackage{feynmp-auto}
\DeclareSymbolFont{forjmath}{OT1}{cmr}{m}{sl}
\DeclareMathSymbol{\Jmath}{\mathord}{forjmath}{'021}
\def\jmath{\Jmath}
\DeclareFontFamily{OT1}{cmr}{}
\DeclareFontFamily{OT1}{cmss}{}
\usepackage{pifont}
\usepackage{fancybox}
\usepackage{parskip}
\usepackage{enumerate}
\usepackage{url}
\biboptions{square,numbers,sort&compress}
\providecommand{\DOI}[1]{\href{http://dx.doi.org/#1}{\doi{#1}}}
\providecommand{\doi}{\texttt{doi:}\begingroup \urlstyle{tt}\Url}
\allowdisplaybreaks
\def\csumb{Dipartimento di Fisica Teorica, Universit\`a di Torino, Italy\\
           INFN, Sezione di Torino, Italy}
\def\support{Work supported by MIUR under contract
    2001023713$\_$006 and by the Research Executive Agency (REA) of the European Union
under the Grant Agreement PITN-GA-2012-316704 (HiggsTools).}
\makeatletter
\def\section{\@startsection{section}{0}{\z@}{5.5ex plus .5ex minus
 1.5ex}{2.3ex plus .2ex}{\large\bf}}
\def\subsection{\@startsection{subsection}{1}{\z@}{3.5ex plus .5ex minus
 1.5ex}{1.3ex plus .2ex}{\normalsize\bf}}
\def\subsubsection{\@startsection{subsubsection}{2}{\z@}{-3.5ex plus
-1ex minus  -.2ex}{2.3ex plus .2ex}{\normalsize\sl}}

\renewcommand{\@makecaption}[2]{%
   \vskip 10pt
   \setbox\@tempboxa\hbox{\small #1: #2}
   \ifdim \wd\@tempboxa >\hsize     
       \small #1: #2\par          
     \else                        
       \hbox to\hsize{\hfil\box\@tempboxa\hfil}
   \fi}
 \def\citenum#1{{\def\@cite##1##2{##1}\cite{#1}}}
\def\citea#1{\@cite{#1}{}}
\newcount\@tempcntc
\def\@citex[#1]#2{\if@filesw\immediate\write\@auxout{\string\citation{#2}}\fi
  \@tempcnta\z@\@tempcntb\m@ne\def\@citea{}\@cite{\@for\@citeb:=#2\do
    {\@ifundefined
       {b@\@citeb}{\@citeo\@tempcntb\m@ne\@citea\def\@citea{,}{\bf }\@warning
       {Citation `\@citeb' on page \thepage \space undefined}}%
    {\setbox\z@\hbox{\global\@tempcntc0\csname b@\@citeb\endcsname\relax}%
     \ifnum\@tempcntc=\z@ \@citeo\@tempcntb\m@ne
       \@citea\def\@citea{,}\hbox{\csname b@\@citeb\endcsname}%
     \else
      \advance\@tempcntb\@ne
      \ifnum\@tempcntb=\@tempcntc
      \else\advance\@tempcntb\m@ne\@citeo
      \@tempcnta\@tempcntc\@tempcntb\@tempcntc\fi\fi}}\@citeo}{#1}}
\def\@citeo{\ifnum\@tempcnta>\@tempcntb\else\@citea\def\@citea{,}%
  \ifnum\@tempcnta=\@tempcntb\the\@tempcnta\else
  {\advance\@tempcnta\@ne\ifnum\@tempcnta=\@tempcntb \else\def\@citea{--}\fi
    \advance\@tempcnta\m@ne\the\@tempcnta\@citea\the\@tempcntb}\fi\fi}
\makeatother
\DeclareRobustCommand{\PA}{\HepParticle{A}{}{}\Xspace}
\DeclareRobustCommand{\PV}{\HepParticle{V}{}{}\Xspace}
\DeclareRobustCommand{\PX}{\HepParticle{X}{}{}\Xspace}

\DeclareRobustCommand{\Pf}{\HepParticle{f}{}{}\Xspace}
\DeclareRobustCommand{\PAf}{\HepAntiParticle{\Pf}{}{}\Xspace}
\DeclareRobustCommand{\PF}{\HepParticle{F}{}{}\Xspace}

\DeclareRobustCommand{\PL}{\HepParticle{L}{}{}\Xspace}

\newcommand{\PAA}{\PA\PA}

\newcommand{\PVV}{\PV\PV}

\newcommand{\SMEFT}{\rm{\scriptscriptstyle{SMEFT}}}

\newcommand{\myLO}{\mathrm{\scriptscriptstyle{LO}}}
\newcommand{\myNLO}{\mathrm{\scriptscriptstyle{NLO}}}

\newcommand{\myQED}{\mathrm{QED}}

\newcommand{\mySM}{\rm{\scriptscriptstyle{SM}}}

\newcommand{\ssB}{{\mathrm{B}}}

\newcommand{\ssQ}{{\mathrm{Q}}}

\newcommand{\ssF}{{\mathrm{F}}}

\newcommand{\ssR}{{\mathrm{R}}}

\newcommand{\ssL}{{\mathrm{L}}}

\newcommand{\ssN}{{\mathrm{N}}}

\newcommand{\mrc}{{\mathrm{c}}}

\newcommand{\bqas}{\begin{eqnarray*}}
\newcommand{\eqas}{\end{eqnarray*}}
\newcommand{\nl}{\nonumber\\}

\newcommand{\lpar}{\left(}                            
\newcommand{\rpar}{\right)}

\newcommand{\bq}{\begin{equation}}                    
\newcommand{\eq}{\end{equation}}
\newcommand{\bqs}{\begin{equation*}}                    
\newcommand{\eqs}{\end{equation*}}
\newcommand{\bqa}{\arraycolsep 0.14em\begin{eqnarray}}
\newcommand{\eqa}{\end{eqnarray}}
\newcommand{\ba}[1]{\begin{array}{#1}}
\newcommand{\ea}{\end{array}}
\newcommand{\ben}{\begin{enumerate}}
\newcommand{\een}{\end{enumerate}}
\newcommand{\bei}{\begin{itemize}}
\newcommand{\eei}{\end{itemize}}
\newcommand{\eqn}[1]{Eq.~(\ref{#1})}

\newcommand{\bmid}{\Bigr|}

\newcommand{\Bref}[1]{Ref.~\cite{#1}}
\newcommand{\Brefs}[1]{Refs.~\cite{#1}}

\newcommand{\eg}{e.g.\xspace}
\newcommand{\ie}{i.e.\xspace}
\newcommand{\etc}{etc.\@\xspace}

\newcommand{\mw}{\mathswitch{M_{\PW}}}

\newcommand{\mz}{\mathswitch{M_{\PZ}}}

\newcommand{\mws}{\mathswitch{M^2_{\PW}}}

\newcommand{\mzs}{\mathswitch{M^2_{\PZ}}}

\newcommand{\mh}{\mathswitch{M_{\PH}}}

\newcommand{\mhc}{\mathswitch{M^2_{\sPH}}}
\newcommand{\mhs}{\mathswitch{M^2_{\sPH}}}

\newcommand{\sPF}{{\scriptscriptstyle{\PF}}}

\newcommand{\sPH}{{\scriptscriptstyle{\PH}}}

\newcommand{\sPZ}{{\scriptscriptstyle{\PZ}}}
\newcommand{\sPW}{{\scriptscriptstyle{\PW}}}
\newcommand{\sPA}{{\scriptscriptstyle{\PA}}}
\newcommand{\sPB}{{\scriptscriptstyle{\PB}}}
\newcommand{\sPV}{{\scriptscriptstyle{\PV}}}

\newcommand{\sPHAA}{{\scriptscriptstyle{\PH\PAA}}}
\newcommand{\sPHAZ}{{\scriptscriptstyle{\PH\PA\PZ}}}
\newcommand{\sPHAV}{{\scriptscriptstyle{\PH\PA\PV}}}

\newcommand{\cph}{\mathswitch{s_{\sPH}}}

\newcommand{\cpz}{\mathswitch{s_{\sPZ}}}

\newcommand{\muh}{\mathswitch{\mu_{\PH}}}

\newcommand{\muhs}{\mathswitch{\mu^2_{\PH}}}

\newcommand{\gh}{\mathswitch{\gamma_{\PH}}}

\newcommand{\shat}{\mathswitch{\hat s}}

\newcommand{\proc}{{\mbox{\scriptsize proc}}}

\newcommand{\eff}{{\mbox{\scriptsize eff}}}

\newcommand{\gen}{{\mbox{\scriptsize gen}}}

\newcommand{\ep}{\mathswitch \varepsilon}

\newcommand{\spro}[2]{{#1}\cdot{#2}}

\newcommand{\lrbr}{\left[}
\newcommand{\rrbr}{\right]}

\newcommand{\mrS}{{\mathrm{S}}}
\newcommand{\mrA}{{\mathrm{A}}}

\newcommand{\mrE}{{\mathrm{E}}}
\newcommand{\mrF}{{\mathrm{F}}}

\newcommand{\mrH}{{\mathrm{H}}}
\newcommand{\mrI}{{\mathrm{I}}}

\newcommand{\mrO}{{\mathrm{O}}}
\newcommand{\mrP}{{\mathrm{P}}}

\newcommand{\mrT}{{\mathrm{T}}}

\newcommand{\Lag}{{\cal L}}

\newcommand{\srt}{\sqrt{2}}
\newcommand{\myGF}{G_{\sPF}}
\DeclareRobustCommand{\PK}{\HepParticle{\upPhi}{}{}\Xspace}

\DeclareRobustCommand{\PKdag}{\HepParticle{\PK}{}{\dagger}\Xspace}

\DeclareRobustCommand{\PAf}{\HepAntiParticle{\Pf}{}{}\Xspace}
\newcommand{\mrs}{\mathrm{s}}
\newcommand{\stw}{\mrs_{_{\theta}}}             
\newcommand{\ctw}{\mrc_{_{\theta}}}
\newcommand{\stws}{\mrs_{_{\theta}}^2}
\newcommand{\ctws}{\mrc_{_{\theta}}^2}

\newcommand{\stwq}{\mrs_{_{\theta}}^4}

\newcommand{\cths}{{\hat \mrc}^2_{_{\theta}}}

\newcommand{\sths}{{\hat \mrs}^2_{_{\theta}}}

\DeclareRobustCommand{\PXpm}{\HepParticle{\PX}{}{\pm}\Xspace}

\newcommand{\fact}{{\mbox{\scriptsize fc}}}
\newcommand{\nfact}{{\mbox{\scriptsize nfc}}}
\newcommand{\spin}{{\mbox{\scriptsize spin}}}

\newcommand{\aAA}{a_{\scriptscriptstyle{\PAA}}}

\newcommand{\apBox}{a_{\upphi\,\scriptscriptstyle{\Box}}}
\newcommand{\apWB}{a_{\upphi\,\scriptscriptstyle{\PW\PB}}}

\newcommand{\apB}{a_{\upphi\,\scriptscriptstyle{\PB}}}
\newcommand{\apW}{a_{\upphi\,\scriptscriptstyle{\PW}}}
\newcommand{\apD}{a_{\upphi\,\scriptscriptstyle{\PD}}}

\newcommand{\atB}{a_{\PQt\,\scriptscriptstyle{\PB}}}
\newcommand{\atW}{a_{\PQt\,\scriptscriptstyle{\PW}}}
\newcommand{\abB}{a_{\PQb\,\scriptscriptstyle{\PB}}}
\newcommand{\abW}{a_{\PQb\,\scriptscriptstyle{\PW}}}
\newcommand{\atlequ}{a^{(3)}_{\Pl\Pe\PQq\PQu}}

\newcommand{\atp}{a_{\PQt\,\upphi}}

\newcommand{\abp}{a_{\PQb\,\upphi}}

\newcommand{\alW}{a_{\Pl\,\scriptscriptstyle{\PW}}}
\newcommand{\alB}{a_{\Pl\,\scriptscriptstyle{\PB}}}

\newcommand{\aplt}{a^{(3)}_{\upphi\,\Pl}}

\newcommand{\apqt}{a^{(3)}_{\upphi\,\PQq}}
\newcommand{\mcA}{\mathcal{A}}

\newcommand{\mcO}{\mathcal{O}}

\newcommand{\mcT}{\mathcal{T}}

\newcommand{\mrdim}{\mathrm{dim}}
\newcommand{\spc}{\,,}
\newcommand{\spp}{\,.}
\newcommand{\gds}{g_{_6}}

\newcommand{\redeq}[1]{{\color{BrickRed}{#1}}}
\newcommand{\proA}[1]{\Delta_{\sPA}(#1)}
\newcommand{\proZ}[1]{\Delta_{\sPZ}(#1)}

\newcommand{\SR}{\mathrm{SR}}
\newcommand{\NR}{\mathrm{NR}}
\newcommand{\sPAZ}{{\scriptscriptstyle{\PA\PZ}}}
\newcommand{\bmidl}{\Bigr|\Xspace}
\newcommand{\bmidr}{\Xspace\Bigr|}
\newcommand{\proZb}[1]{{\overline{\Delta}}_{\sPZ}(#1)}
\newcommand{\muV}{\mathswitch{\mu_{\sPV}}}
\newcommand{\muVs}{\mathswitch{\mu^2_{\sPV}}}
\newcommand{\gamV}{\mathswitch{\gamma_{\sPV}}}
\newcommand{\gamVs}{\mathswitch{\gamma^2_{\sPV}}}
\newcommand{\reg}{{\mbox{\scriptsize reg}}}
\newcommand{\muZ}{\mathswitch{\mu_{\sPZ}}}
\newcommand{\muZs}{\mathswitch{\mu^2_{\sPZ}}}
\newcommand{\gamZ}{\mathswitch{\gamma_{\sPZ}}}

\newcommand{\PO}{\mathrm{\scriptscriptstyle{PO}}}
\DeclareRobustCommand{\PWpL}{\HepParticle{\PW}{\PL}{+}\Xspace}
\DeclareRobustCommand{\PWmL}{\HepParticle{\PW}{\PL}{-}\Xspace}
\DeclareRobustCommand{\PAe}{\HepAntiParticle{\Pe}{}{}\Xspace}
\newcommand{\tcif}{\mathswitch {I^{(3)}_{\Pf}}}
\newcommand{\sft}{{\mbox{\scriptsize soft}}}
\newcommand{\gsix}{\mathswitch {g_{_6}}}
\journal{Reviews in Physics}

\begin{document}

\begin{frontmatter}
\title{Through precision straits to next standard model heights}

\author[cern]{Andr\'{e} David}
\ead{andre.david@cern.ch}

\author[torino]{Giampiero Passarino\fnref{support}}
\ead{giampiero@to.infn.it}

\address[cern]{PH Department, CERN, Switzerland}
\address[torino]{\csumb}

\fntext[support]{\support}

\begin{abstract}

After the LHC Run~1,
the standard model (SM) of particle physics has been completed.
Yet, despite its successes, the SM has shortcomings
vis-\`{a}-vis cosmological and other observations.
At the same time, while the LHC restarts for Run 2 at 13 TeV,
there is presently a lack of direct evidence
for new physics phenomena at the
accelerator energy frontier.
From this state of affairs arises the need for
a consistent theoretical framework
in which deviations from the SM predictions can be calculated
and compared to precision measurements.
Such a framework should be able to comprehensively
make use of all measurements
in all sectors of particle physics,
including
LHC Higgs measurements,
past electroweak precision data,
electric dipole moment,
$g-2$,
penguins and flavor physics,
neutrino scattering,
deep inelastic scattering,
low-energy $e^{+}e^{-}$ scattering,
mass measurements,
and any search for physics beyond the SM.
By simultaneously describing all existing measurements,
this framework then becomes
an intermediate step,
pointing us toward the next SM,
and hopefully revealing the underlying symmetries.
We review the role that
the standard model effective field theory (SMEFT)
could play in this context,
as a consistent, complete, and calculable generalization
of the SM in the absence of light new physics.
We discuss the relationship of the SMEFT
with the existing kappa-framework for Higgs boson couplings
characterization
and the use of pseudo-observables,
that insulate experimental results
from refinements due to ever-improving calculations.
The LHC context, as well as that of previous and future accelerators
and experiments, is also addressed.
\end{abstract}

\end{frontmatter}

\section{The Higgs boson}
During the LHC Run~1
a new resonance was discovered in 2012~\cite{Aad:2012tfa,Chatrchyan:2012ufa}.
That resonance, with a mass of $125\pm0.24\UGeV$~\cite{Aad:2015zhl},
is a candidate to be the Higgs boson of the standard model (SM).
The spin-$0$ nature of the resonance
is well established~\cite{Bolognesi:2012mm,Aad:2013xqa,Khachatryan:2014kca},
all the available studies on the couplings
of the new resonance conclude it to be compatible with
the Higgs boson of the SM
within present precision~\cite{Khachatryan:2014jba,Aad:2015gba},
and, as of yet,
there is no direct evidence for new physics phenomena beyond the SM (BSM).

Inevitably, after the LHC Run~1 results
comes a need for a better understanding of the
current ``we haven't seen anything (yet)''
theoretical \textit{zeitgeist}.
Is the SM with a $125\UGeV$ Higgs boson
the final theory, or indeed can it be?
The associated problems with the SM
are known and include the neutrino masses
as well as cosmological evidence for dark matter.

The discovery of a scalar resonance and the absence of
direct evidence for new physics forces us to change perspectives
and to redefine the problem.
In this review, the starting point is to assume
quantum field theory (QFT) as the framework with which to
study the basic constituents of matter.
The parameters of QFT Lagrangians describe the dynamics,
something that is at the heart of the needed change of perspective.
At LEP, the dynamics were fixed by the SM Lagrangian,
with the unknowns being parameters
such as the Higgs mass $\mh$,
the strong coupling constant $\alphas(\mz)$,
\etc~\cite{SomethingOnParameters}.
In other words,
at LEP the SM was the hypothesis
and bounds on $\mh$ were derived
from a comparison with high-precision data.
At the LHC, after the 2012 discovery,
the unknowns are deviations from the SM,
given that the SM is fully specified
and constrained by experimental measurements
of increasing precision and accuracy.
The definition of SM deviations
requires a characterization of the underlying dynamics.
Whereas (concrete) BSM models represent specific roads toward the Planck scale,
it would be of great interest to employ a (more) model-independent approach,
a framework that could describe a whole class of paths to the Planck scale.

While studies performed with limited precision may only claim the discovery of
a SM-like Higgs boson, as soon as greater precision is available,
it may be possible to decipher the
nature of the Higgs through
the accurate determination of its couplings~\cite{Ellis:2013lra,Englert:2014uua,Cranmer:2013hia,Asner:2013psa}.

Given the precision that was expected for LHC Run~1 results,
it was natural
to begin exploring the couplings using
the (original) $\upkappa\,$-framework~\cite{LHCHiggsCrossSectionWorkingGroup:2012nn,Heinemeyer:2013tqa}.
There is no need to repeat here the main argument,
of splitting and scaling different loop
contributions in the amplitudes of processes mediated by Higgs bosons.
The main shortcoming is that the
original $\upkappa\,$-framework is only an intuitive language
that lacks internal consistency
when moving beyond leading order (LO).
In parallel, recent years have witnessed an increasing
interest in Higgs effective Lagrangians and SM effective field theory (EFT);
see in particular
\Brefs{Contino:2013kra,Azatov:2014jga,Contino:2014aaa},
\Brefs{Berthier:2015oma,Trott:2014dma,Alonso:2013hga,Jenkins:2013wua,Jenkins:2013sda,Jenkins:2013zja,Jenkins:2013fya},
\Brefs{Artoisenet:2013puc,Alloul:2013naa},
\Bref{Ellis:2014dva},
\Bref{Falkowski:2014tna},
\Bref{Low:2009di},
\Brefs{Degrande:2012wf,Chen:2013kfa},
\Bref{Grober:2015cwa},
\Brefs{Englert:2015bwa,Englert:2015zra,Englert:2014uua} and
\Brefs{Biekoetter:2014jwa,Gupta:2014rxa,Elias-Miro:2013mua,Elias-Miro:2013gya,Pomarol:2013zra,Masso:2014xra}.
EFTs can be used to describe the full set of deviations from the SM and therefore
a better name is certainly SMEFT,
as used in \Bref{Henning:2014wua,Alonso:2014rga} and
\Brefs{Hartmann:2015oia,Berthier:2015oma}.

It is worth noting that there is no formulation
which is completely model-independent and
the SMEFT, as any other approach,
is based on a given set of (well-defined) assumptions.
In full generality we can distinguish
a top-down approach (model-dependent) and
a bottom-up approach (with fewer assumptions).
The top-down approach is based on several steps.
First one has to classify BSM models,
possibly respecting custodial symmetry
and decoupling of high mass states,
then the corresponding SMEFT can be
constructed, \eg via a covariant derivative expansion~\cite{Henning:2014wua}.
Once the SMEFT is derived one can construct
the corresponding SM deviations,
that may be different for each BSM model or class of BSM models.
The bottom-up approach starts
with the determination of a basis of $\mrdim = 6$ (or higher) operators
and proceeds directly to the classification of SM deviations,
possibly respecting
the analytic structure of the SM amplitudes.
The synthesis is that $\mrdim= 6$ operators
are supposed to arise from a local Lagrangian,
containing heavy degrees of freedom decoupled
from the presently-probed energy scales.
Of course, the correspondence between Lagrangians and effective operators
is not bijective because different Lagrangians can give rise to the
same operator.

The change of perspective after the LHC Run~1 is equivalent to saying
that we have moved from a fully predictive (SM) phase
to a ``partially predictive (fitting)'' one.
The predictive phase is defined as follows:
in any (strictly) renormalizable theory with $n$
parameters we need to match $n$ data points,
and the $(n+1)$-th calculation is a prediction, \eg
as can be done in the SM.
In the fitting (partially predictive)
phase there will be $(N_6{+}N_8{+}\,\dots = \infty)$
renormalized Wilson coefficients to be fitted, \eg by measuring
the SM deformations due to a single $\mathcal{O}^{(6)}$ insertion.
This represents a departure from the use of a strictly renormalizable theory,
with the compromise of gaining, order-by-order,
the ability to explore deviations
that can only be constrained by fitting to data.
As the number of parameters increases
it becomes inevitable that only combinations
of the parameters can be constrained.

There is a conceptual difference between Higgs physics at the LHC,
for which the UV completion is unknown,
and other scenarios where EFT techniques are applied
and for which there are known UV completions.
When the UV completion is known,
we consider a theory with both light and heavy particles;
the Lagrangian is $\Lag(m)$ where $m$ is the mass of the heavy degree of freedom.
Next, we introduce the corresponding $\Lag_{\eff}$,
the effective theory valid up to a scale $\Lambda = m$.
We renormalize the two theories, say in the $\overline{\rm{MS}}\,$-scheme,
taking care that loop-integration
and heavy limit are operations that do not commute,
and impose matching conditions among renormalized
``light'' one-particle irreducible ($1$PI) Green's functions.

When we compare the present situation with the past an analogy can be drawn.
Consider the QED Lagrangian and complement it with $\mrdim = 6$ Fermi operators
$\PAe_{\ssL}\,\gamma^{\mu}\,\Pe_{\ssL}\,\PAe_{\ssL}\,\gamma_{\mu}\,\Pe_{\ssL}$, \etc.
This EFT can be used to study the muon decay but also
$\PGn_{\Pe}(\PGn_{\PGm}) \Pe$ scattering in the
approximation of zero momentum transfer.
Using data on $\sigma_{\PAGnl\Pe}/\sigma_{\PGnl\Pe}$,
one can derive predictions for the $\PZ$ couplings~\cite{Passarino:1989ey},
\eg for the ratio $g^{\Pe}_{\sPV}/g^{\Pe}_{\sPA}$.
In principle, one could have realized the possibility of having neutral currents.
Understanding that the Yang-Mills theory could match this EFT
at very low energy scales took longer~\cite{Veltman:1968ki}, and
pretending to use this theory to describe the $\PZ\,$-lineshape is not feasible
as the $\PZ$ boson mass is beyond the validity of this EFT.

One could ask: would there be a way to take the Fermi theory
and show how this theory would have pointed to massive vector bosons?
The answer is yes, due to unitarity violations at large
energies (pure $\mathrm{S}\,$-wave unitarity);
for instance, with intermediate vector bosons the
$\PGne \Pem \to \PGne \Pem$ scattering is better behaved
as it is no longer a pure $\mathrm{S}\,$-wave process.
For $\PGnGm \Pem \to \PGne \PGmm$ scattering,
unitarity applied to the $\mathrm{l} = 0$ LO partial wave
requires that $\mrE_{\mathrm{cm}} < (\pi/2\,\sqrt{2}\,\myGF)^{1/2} \approx 310\UGeV$.
Furthermore, the interaction had a well-known structure,
\eg in neutron decay,
muon decay,
and neutrino events,
that strongly suggested the existence of massive spin-$\,1$ particles.
In hindsight, the Fermi Lagrangian could have been built from symmetries
(of the SM) only,
\ie left-handed leptons are doublets under $SU(2)$
and flavor universality.
In that case the Fermi theory contains
the only $\mrdim= 6$ operator with a charged current (CC).

Retrospectively one could have written
\bq
\Lag_{\ssF} = \myGF\,
{\overline{\psi}}\,\psi\;
{\overline{\psi}}\,\psi =
\sum_i\,{\overline{\psi}}_{\Pp}\,\mrO_i\,\psi_{\Pn}\;\Bigl[
C_i\,{\overline{\psi}}_{\Pe}\,\mrO_i\,\psi_{\PGn} +
C'_i\,{\overline{\psi}}_{\Pe}\,\mrO_i\,\gamma^5\,\,\psi_{\PGn} \Bigr] +
\mbox{h. c.} \spc
\eq
where the $\mrO_i$ refer to scalar,
$\dots$, tensor structures, and extended it to become
\bq
\Lag_{\ssF} = \myGF\,
{\overline{\psi}}\,\psi\;
{\overline{\psi}}\,\psi +
a_{\scriptscriptstyle{\Box}}\,\myGF^2\,
{\overline{\psi}}\,\psi\,
\Box\,
{\overline{\psi}}\,\psi + \dots
\eq
add counterterms,
making it possible for the theory to become predictive at the loop level.

Historically, events went differently:
charged currents were measured to be flavor universal,
parity violation was discovered,
the V-A structure detected,
the $SU(2)$ symmetry was postulated,
and neutral currents (NC) predicted;
finally NCs were discovered and the SM made its success.

The SMEFT used so far is based on several assumptions:
one Higgs doublet with a linear representation
(for non-linerar see \Bref{Buchalla:2013rka}),
no new ``light'' degrees of freedom and decoupling of heavy degress of freedom,
and absence of mass mixing of new heavy scalars with the SM Higgs doublet.
Furthermore,
most of the approaches present in the literature are LO SMEFT,
\ie they include SM up to next-to-leading order (NLO)
and SMEFT ``contact'' terms.
There are two directions for improving upon this scenario:
adding $\mrdim > 6$ operators without touching the SM loops and
inserting $\mrdim > 4$ operators in SM loops.
Ideally one should move along the diagonal direction in this space, doing both.

In \Bref{Ghezzi:2015vva} it was re-established that
a SMEFT can provide an adequate answer
for describing SM deviations beyond LO.
The direction chosen in \Bref{Ghezzi:2015vva} and also
in \Brefs{Hartmann:2015oia,Hartmann:2015aia}
is to work with the insertion of $\mrdim = 6$ operators.
In this construction,
the scale $\Lambda$ that characterizes BSM physics cannot be too small,
because $\mrdim = 8$ operators were neglected.
But $\Lambda$ can also not be too large,
because $\mrdim = 4$ higher-order corrections
may be more important than
the $\mrdim = 6$ interference effects.
It is worth noting that these statements
do not imply an inconsistency of SMEFT.
They only mean that higher-dimensional operators and/or
higher-order electroweak (EW) corrections
(\eg \Bref{Actis:2008ts}) must also be included
if one wants to explore larger ranges of $\Lambda$.
The general SMEFT decomposition of any amplitude, projected into the
$\alphas = 0$ plane, reads as follows:
\bq
\mcA = \sum_{n=\ssN}^{\infty}\,g^n\,\mcA^{(4)}_n +
       \sum_{n=\ssN_6}^{\infty}\,\sum_{l=0}^n\,\sum_{k=1}^{\infty}\,
        g^n\,g^l_{4+2\,k}\,\mcA^{(4+2\,k)}_{n\,l\,k} \spc
\eq
where $g$ is the $SU(2)$ coupling constant, and
$g_{4+2\,k} = 1/(\sqrt{2}\,G_{\ssF}\,\Lambda^2)^k$.
For each process, $N$ defines the LO for $\mrdim = 4$:
\eg $N = 1$ for $\PH \to \PV\PV$
and other tree-level couplings,
but $N = 3$ for $\PH \to \PGg\PGg$,
a loop-induced process at LO.
$N_6 = N$ for tree-level processes and
$N_6 = N - 2$ for loop-induced processes.

Generally speaking, there is no factorization of SM higher-order terms.
Therefore, reweighing the leading-order SM predictions
to account for higher-order QCD and EW corrections,
assuming factorization from the EFT effects,
is not a procedure
that produces accurate results (see \Bref{Aad:2015tna}).
As far as QCD factorization is concerned
let us consider the well-known example
\bq
\Pg(p_1) + \Pg(p_2) \to \PA(p_a) + \PB(p_b) + \PX
\quad \lpar p_1 = z x_1 P_1,\; p_2 = x_2 P_2\rpar \spc
\eq
where $\lpar p_a + p_b \rpar^2 = Q^2$,
$\tau s = Q^2$,
and $z \to 1$ is the soft limit
\bq
d\,\sigma\lpar \tau\,,\,Q^2\,,\,\dots \rpar =
\int d x_1\,d x_2\, d z\,
f_{\Pg}\lpar x_1\,,\,\mu_{\ssF} \rpar\,f_{\Pg}\lpar x_2\,,\,\mu_{\ssF} \rpar\,
\delta\lpar \tau - x_1 x_2 z\rpar
d\,{\hat{\sigma}}\lpar z\,,\,\alphas\,,\,
\frac{Q^2}{\mu^2_{\ssR}}\,,\,\frac{Q^2}{\mu^2_{\ssF}}\,\dots \rpar \spc
\eq
where $d\,{\hat{\sigma}} = d\,{\hat{\sigma}^0}\,z\,G$ and
\bq
G^{\myNLO}\lpar z\,,\,\alphas \rpar\bmid_{\sft} = \delta\lpar 1 - z\rpar +
\frac{\alphas}{2\,\pi}\,\Bigl[
d_1\,D_1(z) + \lpar c_0 + c_1 + \dots \rpar\,\delta\lpar 1 - z \rpar \Bigr] \spc
\quad
D_n(z) = \Bigl[ \frac{\ln^n(1-z)}{1-z} \Bigr]_{+} \spp
\eq
Non-universal NLO corrections (that are process-dependent)
enter through the coefficient $c_1$ and $D_n(z)$
(plus sub-leading terms,)
imply convolution, and dominate the cross-section in the soft
limit.
For re-evaluation it is important to have the answer in terms of SM deviations:
this allows to ``reweight'' when new
(differential) $\mathrm{K}\,$-factors become available.
New input will touch only the $\mrdim= 4$ components.

The rationale in building a QFT of SM deviations
is not so much the numerical impact of
higher orders (even if some can be sizable)
but in promoting a phenomenological tool
(the $\upkappa\,$-framework) to the full-fledged status of QFT.
Another reason for having a
complete formalism is to
avoid a situation where experimentalists
will have to go back and ``remove''
a provisional formalism from the analysis.
To explain SMEFT in a nutshell,
consider a process described by a SM amplitude
\bq
\mcA_{\mySM} = \sum_{i=1}^{n}\,\mcA^{(i)}_{\mySM} \spc
\eq
where the $\mcA^{(i)}_{\mySM}$
are gauge-invariant sub-amplitudes.
In general, the same process is given
by a contact term or
a collection of contact terms of $\mrdim = 6$;
for instance,
direct coupling of $\PH$ to $\PV\PV (\PV= \PGg, \PZ, \PW)$.
In order to construct the theory
one has to select a set of higher-dimensional operators and
start the complete procedure of renormalization.
Of course, different sets of operators
can be interchanged
as long as they are closed under renormalization.
It is evident that renormalization is best performed when using the so-called
Warsaw basis, see \Bref{Grzadkowski:2010es}.
Moving from SM to SMEFT we obtain
\bq
\mcA_{\SMEFT} = \sum_{i=1}^{n}\,\upkappa_i\,\mcA^{(i)}_{\mySM} +
                \gds\,\upkappa_{\mrc} +
                \gds\,\sum_{i=1}^{\ssN}\,a_i\,\mcA^{(i)}_{\nfact} \spc
\label{nloeft}
\eq
where $g^{-1}_{_6} = \sqrt{2}\,\myGF\,\Lambda^2$
and $\upkappa_i = 1 + \gds\,\Delta\upkappa_i$.
The last term in \eqn{nloeft} collects
all NLO contributions that do not factorize (nfc) and
the $a_i$ are Wilson coefficients.
The $\Delta\upkappa_i$ are linear combinations of the $a_i$.

\begin{figure}[t]
   \centering

   \includegraphics[width=0.7\textwidth, trim = 30 250 50 80, clip=true]{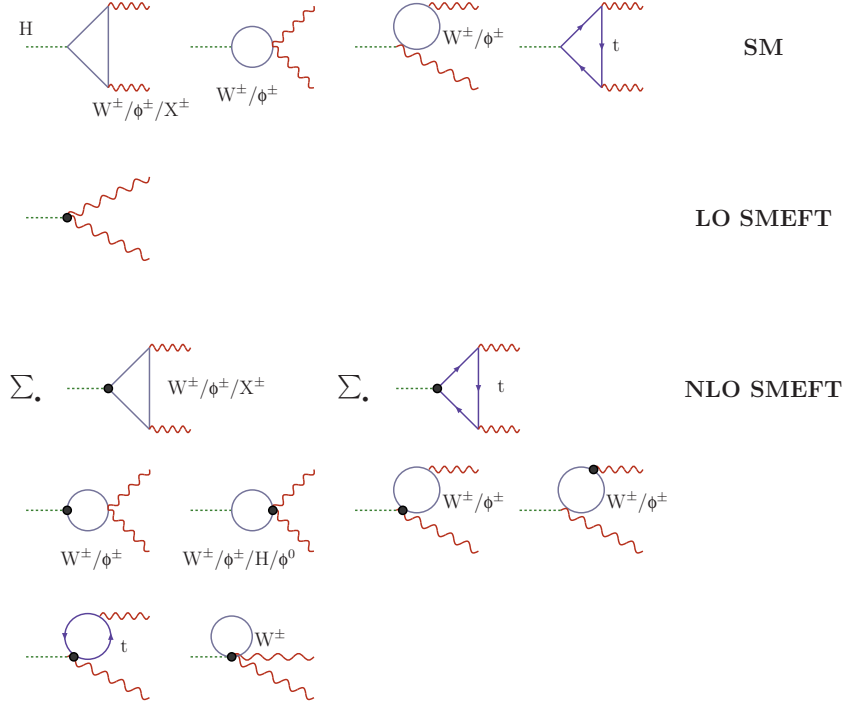}
 \caption{
Diagrams contributing to the amplitude for
$\PH \to \PGg\PGg$ in the $\mathrm{R}_{\xi}\,$-gauge:
SM (first row), LO SMEFT (second row), and NLO SMEFT.
Black circles denote the insertion of one $\mathrm{dim} = 6$ operator.
$\sum_{_{\bullet}}$ implies summing over all insertions
in the diagram (vertex by vertex).
For triangles with internal charge flow
($\PQt, \PW^{\pm}, \upphi^{\pm},\PXpm$)
only the clockwise orientation is shown.
Non-equivalent diagrams obtained
by the exchange of the two photon lines are not shown.
Higgs and photon wave-function factors are not included.
The Fadeev-Popov ghost fields are denoted by $\PX$.
\label{fig:HAA}
}

\end{figure}

We conclude that \eqn{nloeft}
gives the correct generalization of the original
$\upkappa\,$-framework at the price of introducing additional,
non-factorizable, terms in the amplitude.
In strict LO SMEFT and in the linear realization,
only the $\upkappa_{\mrc}$ contact term is included
with the following drawback:
$\upkappa_{\mrc}$ is non-zero
but $\Delta\upkappa_i = 0$.
Therefore, when measuring a deviation from the SM prediction
we would find a non-zero value for $\upkappa_{\mrc}$.
However, at NLO
the $\Delta\upkappa_i$ are non-zero,
leading to a (NLO) degeneracy.
The interpretation
in terms of $\upkappa^{\myLO}_{\mrc}$ or
in terms of $\{\upkappa^{\myNLO}_{\mrc}\,,\,\Delta\upkappa^{\myNLO}_i\}$
is rather different.
Indeed,
mapping of experimental constraints to Wilson coefficients at LO, or at NLO,
should be corrected for
if an inferred constraint on a coefficient
is to be used in predicting another process.
For the $\PH \to \PGg\PGg$ decay process,
within LO SMEFT we ``measure" just
 $\aAA = \stws\,\apW + \ctws\,\apB + \stw \ctw\,\apWB$,
 while at NLO there are contributions proportional to
 $\atW, \atB, \abW, \abB, \apW, \apB, \apWB$
 (with a mixing among $\{\apW, \apB, \apWB\}$).
 The differences are illustrated in Fig.~\ref{fig:HAA}.

In NLO SMEFT each $\upkappa\,$-parameter
has a second index which specifies the corresponding process.
One easily discovers that
there are correlations among the different $\upkappa\,$-parameters
and cross-constraints as well:
this can be seen by solving the inversion problem ($\ctws= \mws/\mzs$):
\bq
\Delta \upkappa^{\sPHAZ}_{\PQb} - \Delta \upkappa^{\sPHAZ}_{\PQt} -
\Delta \upkappa^{\sPHAA}_{\PQb} + \Delta \upkappa^{\sPHAA}_{\PQt} =
\ctws\,\Delta \upkappa^{\sPHAZ}_{\sPW} +
\lpar \frac{3}{2} + 2\,\ctws \rpar\,\lpar \Delta \upkappa^{\sPHAZ}_{\PQt} -
\Delta \upkappa^{\sPHAA}_{\PQt} \rpar +
\lpar \frac{1}{2} + 3\,\ctws \rpar\,\Delta \upkappa^{\sPHAA}_{\sPW} = 0 \spp
\eq
\bqa
\atp &=& \frac{1}{2 \stws}\,\apD - 2\,\apBox + \redeq{\Delta \upkappa^{\sPHAA}_{\PQt}} \spc
\qquad
\abp =  - \frac{1}{2 \stws}\,\apD +
            2\,\apBox - \redeq{\Delta \upkappa^{\sPHAA}_{\PQb}} \spc
\nl
\apBox &=& \frac{1}{4 \stws}\,\apD +
                      \frac{1}{2}\,\redeq{\Delta \upkappa^{\sPHAA}_{\sPW}}
\qquad
2\,\ctws\,\apD = \stws\,\lpar \redeq{\Delta \upkappa^{\sPHAZ}_{\PQb}} -
                     \redeq{\Delta \upkappa^{\sPHAA}_{\PQb}} \rpar \spp
\eqa
Considering decay processes, $\PH \to \{\PF\}$, we define ``effective'' kappas
\bq
\mcA^{\PH\{\PF\}}_{\SMEFT}(s) = \upkappa^{\PH\{\PF\}}(s)\,\mcA_{\mySM}(s)
\qquad
\Gamma_{\sPH} = \sum_{\{\PF\}}\,\mid \upkappa^{\PH\{\PF\}}(\mhs) \mid^2\,
\Gamma^{\mySM}_{\PH \to \{\PF\}}
\spp
\label{gammaH}
\eq
In writing \eqn{gammaH} we have assumed that the Higgs may not decay
to new invisible or undetectable particles.
Another important point to mention is the dependence
of the effective kappas on the scale relevant for the process;
that has the consequence that rescaling couplings at the $\PH$ peak
is not the same thing as rescaling them
off-peak~\cite{Ghezzi:2014qpa,Englert:2015bwa}.
Therefore, off-shell measurements are (much) more
than consistency checks on $\Gamma_{\sPH}$:
observing an excess in the off-shell measurement
will be a manifestation of BSM physics,
which might or might not need
to be in relation with the $\PH$ width.

It is worth noting that (a priori) discarding subsets of $\mrdim = 6$ operators
is not advisable and, as usual,
approximations should be the last step in the procedure,
after full calculations are performed.
The theory of SM deviations,
workable to all orders,
is still in its infancy but
clearly marks the irrelevance of protracted discussion
of which SMEFT basis to use;
a basis is \emph{by definition} closed under renormalization,
and anything that is not a basis,
such as many effective Lagrangians,
should be viewed with due care.
With NLO SMEFT we can study Higgs couplings to very high accuracy
and try to understand sources of deviations that may appear in the data
from multiple sectors.
Potentially, there will be a blurred arrow
in the space of Wilson coefficients pointing
the way to the UV completion of the SM,
and we should simply focus the arrow.

Another important point to mention
is the limit where SM deviations are set to zero.
In this limit one should recover the most accurate SM predictions.
Therefore, it is certainly allowed
to decompose an amplitude
into form factors multiplying Lorentz structures
(as long as the form factors have a known analytical structure)
but the limit where ``anomalous couplings''
vanish (see Eq.~4 of \Bref{Khachatryan:2015mma})
should not be interpreted as the LO value of the SM amplitude.
This is to say that a measurement
that deviated from the LO SM predictions
is not necessarily ``anomalous''.
From a constructive point of view,
it seems reasonable to require a common language in describing SM deviations,
\ie ``$\PH\PVV$ anomalous couplings'' can be easily
incorporated into the more universal notion of SMEFT.

Before the LHC, when there was no measurement of the Higgs boson mass,
there were two interesting scenarios in
$\PV_{\ssL} \PV_{\ssL} \to \PV_{\ssL} \PV_{\ssL}$ longitudinal scattering:
$\mws, \mzs \muchless \mhs \muchless s$
and $\mws, \mzs \muchless s \muchless \mhs$.
With a $125\UGeV$ Higgs boson we analyze a new option,
$\mws, \mzs, \mhs \muchless s$, for which the SM result
for the transition matrix is
\bq
\frac{d}{d t}\,\sigma_{\PV_{\ssL} \PV_{\ssL} \to \PV_{\ssL} \PV_{\ssL}} =
\frac{\bmid \mrT(s,t)\bmid^2}{16\,\pi\,s^2} \spc
\quad
\mrT^0_{\myLO} = \frac{1}{16\,\pi\,s}\,\int_{-s}^0\,dt\,\mrT_{\myLO} \spc
\quad
\mrT^0_{\myLO}\lpar \PWpL \PWmL \to \PWpL \PWmL\rpar \stackrel{s \to \infty}{\sim}
- \frac{\myGF \mhs}{4\,\srt\,\pi} \spp
\eq
As well know, anomalous couplings violate perturbative unitarity.
However, one has to be careful in formulating the problem,
as the region of interest is
$\mws, \mzs, \mhs \muchless s \muchless \Lambda^2$.
In this case,
when $s$ approaches $\Lambda^2$,
the SMEFT must be replaced by its
UV completion
and it makes no sense to study the limit $s \to \infty$ in SMEFT.
However, it is known that heavy degrees of freedom
may induce effects of
{\em delayed} unitarity cancellation~\cite{Ahn:1988fx}
in the intermediate region.
These effects could be detectable in vector boson scattering
if there is enough room between $\mh$ and the scale of BSM physics.
We derive
\bq
\mrT^0_{\mySM + \SMEFT} \sim \sum_{n=0}^2\,\mrT_n\,\lpar G_{\ssF}\,s \rpar^n \spp
\eq
As expected, the SM part contributes to the constant term,
while $\mrdim = 6$ operators have positive powers of $s$
(up to a power of two).
The leading behavior is controlled
by the $\mcO_{\upphi\,\sPW\sPB} = \PKdag\,\tau^a\,\PK\,
\ssF^{\mu\nu}_a\,\ssF^0_{\mu\nu}$ operator.
Delayed unitarity cancellations might very well be
the best window for detecting BSM physics.

\section{Not just the LHC Higgs}
In our quest for a UV completion of the SM
we cannot neglect the sensitivity of electroweak
precision data (EWPD).
By its general nature,
the SMEFT is not confined to describe Higgs couplings
and their SM deviations:
it can be used to reformulate the constraints from EWPD
and to analyze the whole set of processes measurable at LHC and
future colliders,
such as single and multiple gauge boson production,
Drell-Yan physics,
associated production of gauge bosons and jets,
triple gauge coupling searches,
$\mw$,
asymmetries such as $\mrA_{\ssF\ssB}$,
extraction of $\sin^2\theta_{\sPW}$,
\etc
Here we present a few examples of EWPD evaluated in
NLO SMEFT.

\subsection{\texorpdfstring{$\alpha_{\myQED}$ at the mass of the $\PZ$}
{alpha\_QED at the mass of the Z}}
If we neglect loop-generated (LG) operators~\cite{Einhorn:2013kja} in loops,
the following result holds for vacuum polarization:
\bq
\Pi^{(\mrdim=6)}_{\PA\PA}(0) = - 8\,(\ctws/\stws)\,\apD\,\Pi^{(\mrdim=4)}_{\PA\PA}(0) \spp
\eq
One of the key ingredients in computing precision (pseudo-)observables is
$\alpha_{\myQED}$ at the mass of the $\PZ$,
defined by
\bq
\alpha(\mz) = \frac{\alpha(0)}{1
- \Delta \alpha^{(5)}(\mz)
- \Delta \alpha_{\PQt}(\mz)
- \Delta \alpha^{\alpha\alphas}_{\PQt}(\mz)} \spc
\label{ralpha}
\eq
\bq
\Delta \alpha^{(5)}(\mz) =
\Delta \alpha_{\Pl}(\mz) + \Delta \alpha^{(5)}_{\had}(\mz) \spp
\eq
The numerical impact of the different corrections is
\[
\begin{array}{lcl}
\Delta \alpha^{(5)}_{\had}(\mz) & = & 0.0280398 \\
10^4\,\times\,\Delta \alpha_{\Pl}(\mz) & = & 0.0314976 \\
10^4\,\times\,\Delta \alpha_{\PQt}(\mz) & \approx & [ - 0.62\,,\, - 0.55 ] \\
10^4\,\times\,\Delta \alpha^{\alpha\alphas}_{\PQt}(\mz) & \approx &
[ - 0.114\,,\, - 0.095 ] \\
\end{array}
\]
The effect of the SMEFT
is equivalent to multiply $\Delta \alpha_{\Pl,\PQt}(\mz)$
by $1 - \upkappa_{\alpha}$, where
\bq
\upkappa_{\alpha} = 8\,\gsix\,(\ctws/\stws)\,\apD = 0.188\,\apD
\qquad \mbox{for}\;\;\Lambda = 3\UTeV \spp
\eq
Therefore,
$\mid \upkappa_{\alpha}\,\Delta \alpha_{\PQt} \mid > \Delta \alpha_{\Pl}$
and
$\mid \upkappa_{\alpha}\,\Delta \alpha_{\PQt} \mid \approx
\mid \Delta \alpha^{\alpha\alphas}_{\PQt} \mid$.

\subsection{The \texorpdfstring{$\rho\,$}{rho}-parameter}
Consider the following decomposition of the gauge-boson
self-energies (see \Bref{Bardin:1999ak}):
\bqa
S_{\sPW\sPW} = \frac{g^2}{16\,\pi^2}\,\Sigma_{\sPW\sPW} \spc
&\qquad&
S_{\sPZ\sPZ} = \frac{g^2}{16\,\pi^2\,\ctws}\,\lpar
\Sigma_{33} - 2\,\stws\,\Sigma_{3Q} - \stwq\,\Pi_{\sPA\sPA}\,s\rpar \spc
\nl
\Sigma_{\ssF} &=& \Sigma_{\sPW\sPW}(0) - \Re\,\Sigma_{33}(M^2_{\sPZ})
+ \Re\,\Sigma_{3Q}(M^2_{\sPZ})
\eqa
and define $\rho^{-1} = 1 + \frac{\myGF}{2\,\srt\,\pi^2}\,\Sigma_{\ssF} = 0.99490\;$;
$\Delta\rho$ depends on $\apD$,
$\apBox$, $a_{\phi \Pf}$,
and $a^{(1,3)}_{\phi \Pf}$ (with $\Pf = \Pl,\;\PQu,\;\PQd$,)
when considering only
potentially-tree-generated (PTG) operators~\cite{Einhorn:2013kja}.
The leading term,
that should not be used for accurate predictions, is
\bq
\Delta \rho = M^2_{\PQt}\,\Bigl[ \upkappa_{\rho}\,\Delta \rho^{(4)} +
\gsix\,\sum_i\,\Delta \rho^{(6)}_i\,a_i \Bigr] \spc
\quad
\upkappa_{\rho} = 1 + \frac{\gsix}{11}\,\Bigl[ \frac{7}{6}\,\apD +
28\,( a^{(1)}_{\upphi \PQq} + a^{(3)}_{\upphi \PQq}) - 20\,a_{\upphi \PQt}\Bigr] \spc
\eq
where $a_i = \apD,\;a_{\upphi \PQt},\;a^{(1,3)}_{\upphi \PQq}$.
The explicit form for $\Delta \rho^{(6)}_i$ will not be reported here.

\subsection{The \texorpdfstring{$\PW$}{W} mass}
Working in the $\alpha\,$-scheme we can predict $\MW$~\cite{Bardin:1999ak}.
The solution is
\bqa
\frac{\mws}{\mzs} &=& \cths + \frac{\alpha}{\pi}\,\Re\,\Bigl\{
\lpar 1 - \frac{1}{2}\,\gsix\,\apD\rpar\,\Delta^{(4)}_{\PB}(\mw) +
 \sum_{\gen}\,\Bigl[
\lpar 1 + 4\,\gsix\,\aplt\rpar\,\Delta^{(4)}_{\Pl}(\mw)
\nl
{}&+& \lpar 1 + 4\,\gsix\apqt\rpar\,\Delta^{(4)}_{\PQq}(\mw) \Bigr] + \gsix\,\Bigl[
\Delta^{(6)}_{\PB}(\mw) +
\sum_{\gen}\,\lpar
\Delta^{(6)}_{\Pl}(\mw) + \Delta^{(6)}_{\PQq}(\mw) \rpar \Bigr]\Bigr\} \spc
\eqa
where $\Delta^{(4,6)}_i$ (with $i = \Pl, \PQq$, and $\PB$)
are the $\mrdim = 4,6$
corrections due to leptons, quarks, and bosons.
Furthermore, we have introduced
the LO solution (in the $\alpha\,$-scheme)
for the weak-mixing angle:
\bq
\sths = \frac{1}{2}\,\Bigl[ 1 - \sqrt{1 - 4\,\frac{\pi\,\alpha}{\srt\,\myGF\,\mzs}}\Bigr] \spp
\eq
The expansion can be improved
when working within the SM ($\mrdim = 4$),
\eg by expanding in powers of $\alpha(\mz)$.

\subsection{Dijet data}
In the Warsaw basis~\cite{Grzadkowski:2010es} there are
two distinct sets of $\mrdim= 6$ operators:
$\mrdim= 6$ four-fermion operators
(Tab.~3 in \Bref{Grzadkowski:2010es})
and other
$\mrdim = 6$ operators
(Tab.~2 in \Bref{Grzadkowski:2010es})
.
The first set is relevant for
a) NLO SMEFT predictions
involving processes with external fermions
(\eg $\PH \to \PAQb \PQb$, $\PZ \to \PAf \Pf$ \etc)
and for
b) processes dominated by QCD interactions,
such as dijet distributions, \etc
In the first case,
four-fermion operators modify the fermion self-energy,
contributing to the fermion mass renormalization
and to the fermion wave-function factor,
and any $\upPhi \PAf \Pf$ vertex ($\upPhi = \PH, \PZ$ and $\PW$).
Alternatively,
this set of operators is relevant in probing the SM with dijets,
\eg in the study of angular distributions
of dijets in the process $\Pp\Pp \to \mathrm{jj}$,
see \Bref{Domenech:2012ai}.
The relevant partonic processes are $\PQu\PQu \to \PQu\PQu$,
$\PQd\PQd \to \PQd\PQd$, and $\PQu\PQd \to \PQu\PQd$.
It is worth noting that operators such as
$({\overline{\PQq}}_{\ssL}\,\gamma^{\mu}\,\PQq_{\ssL})^2$ are PTG
and their Wilson coefficients
are not necessarily suppressed by the loop factor $1/(16\,\pi^2)$,
which might be important when considering strongly-coupled BSM physics.
At the moment, NLO SMEFT predictions for dijet production are not available.

\subsection{Flavour physics}
Searches for BSM physics in flavour observables
have been interpreted in terms of an effective Hamiltonian
description using $\mrdim=6$ operators \cite{Altmannshofer:2013foa}.
The usage of SMEFT,
instead of an effective Lagragian or Hamiltonian,
would allow to have a consistent and comprehensive
way to also incorporate any observed deviations from the SM
in flavour physics observables.

\subsection{Lepton dipole operators}
The Wilson coefficients $\alW, \alB$, and $\atlequ$
(see Tab.~2 and Tab.~3 of \Bref{Grzadkowski:2010es})
give a LO contribution to
remarkably clean windows to BSM physics,
namely
the $\PGm \to \Pe + \PGg$ decay,
the anomalous magnetic moment of the muon,
and to the electric dipole moment of the electron
(see \Bref{Alonso:2013hga})
.
The NLO calculation is presently not available
and will be useful for future precision studies.

\subsection{Other examples}
Other processes that can be treated within a SMEFT framework are:
top pair production~\cite{Degrande:2010kt,Degrande:2011rt},
neutral triple gauge boson interactions~\cite{Degrande:2013kka},
Higgs boson plus jet production~\cite{Dawson:2014ora},
and
boosted Higgs boson production~\cite{Dawson:2015gka}.

\subsection{Electroweak precision data}
There are several ways to incorporate EWPD.
So far, the most common option has been to
reduce (a priori) the number of $\mrdim = 6$ operators considered.
Open questions regarding this procedure are~\cite{deBlas:2014ula}:
should one fit one $\upkappa$ at a time?
Should one fit first to the EWPD and then to
$\PH$ observables?
A combination of both?
The SMEFT is the framework and we are just at the beginning
of a new phase that should witness
the consolidation of a ``common language''
between the theory and experimental communities,
linking together many different LHC and non-LHC analyses.
In any case, it is essential that the derivation of
constraints is done in a consistent
and basis-independent manner~\cite{Trott:2014dma}.
A recent analysis~\cite{Berthier:2015gja}
reaches conclusions that differ from the usual claim,
namely that it is not justified
to set individual Wilson coefficients
to zero in the analysis of LHC data
as an attempt to incorporate pre-LHC (EWPD) data.

\section{The strategy: from the multi-pole expansion to pseudo-observables to kappas}
Concerning the $\upkappa\,$-framework,
we can say that the $\upkappa\,$-parameters are easy
to understand in terms of
how they change cross sections and partial decay widths.
Extending the framework should be seen
as expressing the $\upkappa$ parameters
in terms of SMEFT coefficients.
One question that remains to be answered is the following:
could we use and translate part of the LEP language,
\eg that of pseudo-observables (PO),
to recast SMEFT parameters into inclusive POs?

What are POs?
To be concise we could say
that what the experimenters do is to collapse
(and/or transform) some ``primordial quantities''
(such as the number of observed events in some pre-defined set-up)
into some ``secondary quantities''
which we feel that are closer to the theoretical
description of the phenomena.
How were POs defined at LEP?
We will give one example:
within the context of the SM,
fiducial observables (FO) at LEP
are described in terms of some set of amplitudes and cross sections:
\bq
A_{\mySM} = A_{\PGg} + A_{\sPZ} + \mbox{non-fact.} \spc
\quad
\sigma\lpar \shat\rpar = \int dz \mrH_{\mathrm{in}} \lpar z,\shat\rpar\,
\mrH_{\mathrm{fin}}\lpar z,\shat\rpar\,{\hat\sigma}\lpar z,\shat \rpar \spc
\eq
where $\mrH_{\mathrm{in, fin}}$ are QED/QCD radiators.
Once the amplitude,
dressed by the weak loop corrections,
is given
we use the fact that in the
SM there are several effects,
such as
the imaginary parts or the $\PGg{-}\PZ$ interference or the pure QED background,
that have a negligible influence on the line shape.
Therefore, POs are determined by fitting FOs
but some ingredients are still
taken from the SM,
making the model-independent results
dependent upon the SM prediction.

In this way, the exact (de-convoluted) cross-section
is successively reduced to a $\PZ$-resonance.
It is a modification of
a pure Breit-Wigner resonance
because of the $s\,$-dependent width:
\bq
\sigma_{\PAf\Pf}\lpar s\rpar = \sigma^{\PAf\Pf}_0\,
\frac{s^2\,\Gamma^2_{\sPZ}}{\lpar s- M^2_{\sPZ}\rpar^2
+ s^2\,\Gamma^2_{\sPZ}/M^2_{\sPZ}}
\qquad
\sigma^{\PAf\Pf}_0 = \frac{12\,\pi}{M^2_{\sPZ}}\,
\frac{\Gamma_{\Pe}\,\Gamma_{\Pf}}{\Gamma^2_{\sPZ}} \spp
\eq
The partial widths are computed
by including all we know about loop corrections.
One needs to specify $\mz$ and
the (remaining) relevant SM parameters for the
SM-complement.
For instance, the explicit formulae for the $\PZ \PAf \Pf$ vertex are
\bq
\rho^{\Pf}_{\sPZ}\,\gamma^{\mu}\,\lrbr \lpar \tcif + i\,a_{\ssL}\rpar \gamma_+
- 2\,Q_{\Pf}\,\upkappa^{\Pf}_{\sPZ}\,\sin^2\theta + i\,a_{\ssQ}\rrbr
= \gamma^{\mu}\,\bigl(\mathcal{G}^{\Pf}_{\sPV} + \mathcal{G}^{\Pf}_{\sPA}\,\gamma^5\bigr) \spc
\eq
where $\gamma_+ = 1 + \gamma^5$, and $a_{\ssQ,\ssL}$ are the SM imaginary
parts.
By definition, the total and partial widths of the $\PZ$ boson include
also QED and QCD corrections.

From LEP to LHC, does history repeat itself?
Why should it?
It should because POs are a platform
between realistic observables
and theory parameters,
allowing experimentalists and theorists to meet half way between;
\ie theorists do not have to run full simulation and reconstruction
and experimentalists do not need to fully unfold
to model-dependent parameter spaces.

Clearly, the LHC is not LEP and there are many differences.
As a consequence, we face new problems,
\eg off-shell LHC physics is not simple
and resonant/non-resonant
are perfectly tied together,
posing severe questions of gauge invariance.

Despite inherent, albeit technical, difficulties
the next job for the LHC is the high-precision study of
SM-deviations;
this will require several steps.
For each process,
one can write down the SMEFT amplitude,
both for resonant and non-resonant parts and
compute fiducial observables.
Then, express the resonant part
as a function of POs without altering the total,
something different from the strategy adopted at LEP.
The SM non-resonant part also changes
and cannot be subtracted.
At this point, conventionally-defined POs can be fit to data,
and later interpreted in terms
of SMEFT Wilson coefficients
(or BSM Lagrangian parameters).

In order to define POs at the LHC
we need various ingredients~\cite{Passarino:2012cb},
\eg multi-pole expansion (MPE), see \Bref{Gonzalez-Alonso:2015bha},
and phase-space factorization.
In any process,
the residues of the poles (starting from maximal degree) are
gauge invariant quantities, see \Bref{Grassi:2001bz}.
The non-resonant part of the amplitude is a gauge-invariant,
multivariate, function.
That is to say that
the residue of the resonant poles can be POs by themselves and
expressing them in terms of other objects (\eg SMEFT Wilson coefficients)
is an operation the can be postponed to an interpretation step.
The end of the chain, when no poles are left,
requires an (almost) model-independent SMEFT
or model-dependent BSM description;
numerically speaking, it depends on the sensitivity of the
measurements to the non-resonant part.

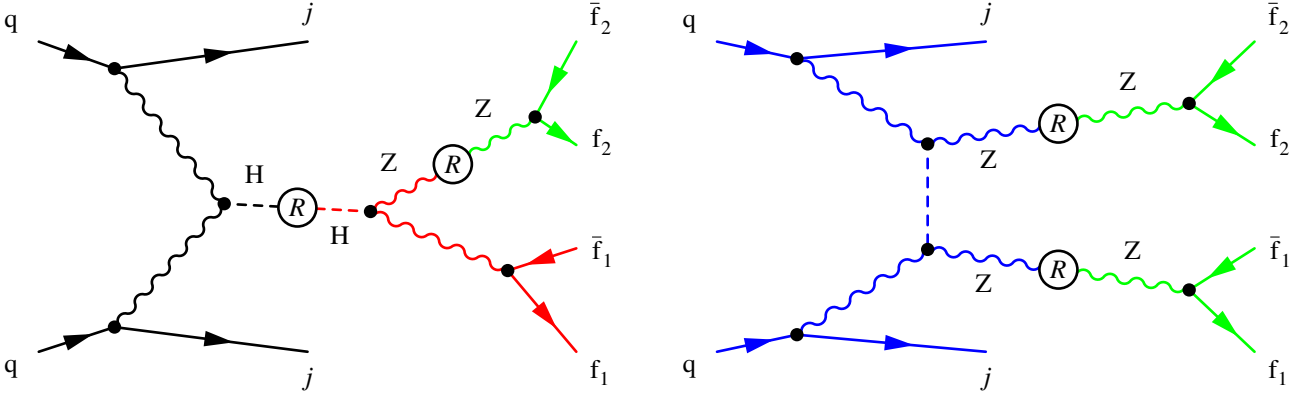
\begin{figure}[t]
  \centering

  \unitlength = \linewidth
\begin{fmffile}{qqffffjj-resonant}
\begin{fmfgraph*}(0.43,0.25)
  \fmfstraight
  \fmfleft{q1,q2}
  \fmfright{f1,f2,f3,f4}
  \fmftop{j2}
  \fmfbottom{j1}

  \fmflabel{\PQq}{q1}
  \fmflabel{\PQq}{q2}
  \fmflabel{$j$}{j1}
  \fmflabel{$j$}{j2}
  \fmflabel{$\Pf_1$}{f1}
  \fmflabel{$\PAf_1$}{f2}
  \fmflabel{$\Pf_2$}{f3}
  \fmflabel{$\PAf_2$}{f4}

  \fmf{fermion}{q1,qjZ1,j1}
  \fmf{fermion}{q2,qjZ2,j2}

  \fmf{phantom, tension=3}{q1,qjZ1}
  \fmf{phantom, tension=3}{q2,qjZ2}

  \fmf{phantom}{ZZH,HZZ}

  \fmf{boson}{ZZH,qjZ1}
  \fmf{boson}{qjZ2,ZZH}
  \fmf{dashes,label=$\PH$}{h2h,ZZH}

  \fmf{dashes, fore=red, label=$\PH$}{h2h,HZZ}
  \fmf{boson, fore=red}{Zff1,HZZ}
  \fmf{boson, fore=red, label=$\PZ$}{HZZ,z2z}
  \fmf{fermion, fore=red}{f2,Zff1,f1}

  \fmf{boson, label=$\PZ$, fore=green}{Zff2,z2z}
  \fmf{fermion, fore=green}{f4,Zff2,f3}

  \fmfv{dec.shape=circle, dec.fill=empty, lab=$R$, lab.dist=0}{h2h,z2z}
  \fmfdot{qjZ1,qjZ2,ZZH,HZZ,Zff1,Zff2}

\end{fmfgraph*}
\end{fmffile}
\hspace{15mm}
%
\begin{fmffile}{qqffffjj-nonresonant}
\begin{fmfgraph*}(0.43,0.25)
  \fmfstraight
  \fmfleft{q1,q2}
  \fmfright{f1,f2,f3,f4}
  \fmftop{j2}
  \fmfbottom{j1}

  \fmflabel{\PQq}{q1}
  \fmflabel{\PQq}{q2}
  \fmflabel{$j$}{j1}
  \fmflabel{$j$}{j2}
  \fmflabel{$\Pf_1$}{f1}
  \fmflabel{$\PAf_1$}{f2}
  \fmflabel{$\Pf_2$}{f3}
  \fmflabel{$\PAf_2$}{f4}

  \fmf{fermion, fore=blue}{q1,qjZ1,j1}
  \fmf{fermion, fore=blue}{q2,qjZ2,j2}

  \fmf{phantom, tension=3}{q1,qjZ1}
  \fmf{phantom, tension=3}{q2,qjZ2}


  \fmf{boson, fore=blue}{HZZ1,qjZ1}
  \fmf{boson, fore=blue}{HZZ2,qjZ2}
  \fmf{dashes, fore=blue}{HZZ1,HZZ2}
  \fmf{boson, fore=blue, label=$\PZ$}{HZZ1,z12z1}
  \fmf{boson, fore=blue, label=$\PZ$}{z22z2,HZZ2}

  \fmf{boson, label=$\PZ$, fore=green}{Zff1,z12z1}
  \fmf{fermion, fore=green}{f2,Zff1,f1}
  \fmf{boson, label=$\PZ$, fore=green}{Zff2,z22z2}
  \fmf{fermion, fore=green}{f4,Zff2,f3}

  \fmfv{dec.shape=circle, dec.fill=empty, lab=$R$, lab.dist=0}{z12z1,z22z2}
  \fmfdot{qjZ1,qjZ2,HZZ1,HZZ2,Zff1,Zff2}

\end{fmfgraph*}
\end{fmffile}
\vspace{5mm}
  \caption{Pseudo-Observables for the triple-resonant (left)
  and double-resonant (right)
  parts of the
  $\pmb{\PQq\PQq \to \PAf_1\Pf_1\PAf_2\Pf_2 jj}$ process at NLO.
  Each color defines one PO and
  the $R$ blobs correspond
  to the cuts induced by the (resonant) $\delta\,$-part of the propagator,
  shown in \eqn{fact}.
  \label{Figcut}}
\end{figure}

The MPE has a dual role:
as we mentioned, poles and their residues are intimately related to the
gauge-invariant splitting of the amplitude (Nielsen identities);
residues of poles (eventually after integration over other variables)
can be interpreted as POs,
something that requires factorization of the amplitude squared.
However, gauge-invariant splitting is not the same as
``factorization'' of the process into sub-processes;
indeed phase-space factorization
requires the pole to be inside the physical region
\bqa
{}&{}&|\Delta|^2 = \frac{1}{\lpar s - M^2\rpar^2 + \Gamma^2\,M^2} =
\frac{\pi}{M\,\Gamma}\,\delta\lpar s - M^2\rpar +
\mathrm{PV}\,\left[ \frac{1}{\lpar s - M^2\rpar^2}\right] \spc
\nl
{}&{}& d\Phi_n\lpar P, p_1 \dots p_n\rpar =
\frac{1}{2\,\pi}\,dQ^2\,d\Phi_{n-(j+1)}\lpar P, Q, p_{j+1} \dots p_n\rpar\,
d\Phi_j\lpar Q, p_1 \dots p_j \rpar \spp
\label{fact}
\eqa
To ``complete'' the decay ($d\Phi_j$) we need
the $\delta\,$-function in \eqn{fact}.
We can say that the $\delta\,$-part of the
resonant propagator opens the corresponding line
and allows us to define POs.
This is not the case for $t\,$-channel propagators,
which cannot be cut.
Consider the process
$\PQq\PQq \to \PAf_1\Pf_1\PAf_2\Pf_2 jj$:
given the structure of the resonant poles we
can define different POs, \eg
\bqa
\sigma(\PQq\PQq \to \PAf_1\Pf_1\PAf_2\Pf_2 jj) &\stackrel{PO}{\longmapsto}&
\sigma(\PQq\PQq \to \PH jj)\,\mathrm{Br}(\PH \to \PZ\PAf_1\Pf_1)\,
\mathrm{Br}(\PZ \to \PAf_2\Pf_2) \spc
\nl
\sigma(\PQq\PQq \to \PAf_1\Pf_1\PAf_2\Pf_2 jj) &\stackrel{PO}{\longmapsto}&
\sigma(\PQq\PQq \to \PZ\PZ jj)\,\mathrm{Br}(\PZ \to \PAf_1\Pf_1)\,
\mathrm{Br}(\PZ \to \PAf_2\Pf_2) \spp
\eqa
These two possibilities are illustrated in Fig.~\ref{Figcut}.
There are fine points to be considered
when factorizing a process into ``physical'' sub-processes.
Consider an amplitude that can be factorized as follows:
\bq
\mcA = \mcA^{(1)}_{\mu}\,\Delta_{\mu\nu}(p)\,\mcA^{(2)}_{\nu} \spc
\eq
where $\Delta_{\mu\nu}$ is the propagator for a spin-$\,1$ resonance.
We would like to replace
\bq
\Delta_{\mu\nu} \to \frac{1}{s - s_{\mrc}}\,\sum_{\lambda}\,
\ep_{\mu}(p,\lambda)\,\ep^*_{\nu}(p,\lambda) \spc
\eq
where $s_{\mrc}$ is the complex pole and $\ep_{\mu}$ are
the spin-$\,1$ polarization vectors.
What we obtain is
\bq
\mid \mcA \mid^2 = \frac{1}{\mid s - s_{\mrc} \mid^2}\,\bmid
\Bigl[ \mcA^{(1)}\,\cdot\,\ep \Bigr]\,\Bigl[ \mcA^{(2)}\,\cdot\,\ep^* \Bigr]\,\bmid^2
\eq
However, extracting the $\delta$ from the propagator does not necessarily factorize the
phase space, \ie we do not find back the needed form,
\bq
\sum_{\lambda} \bmid \mcA^{(1)}\,\cdot\,\ep(p,\lambda) \,\bmid^2\,
\sum_{\sigma}  \bmid \mcA^{(2)}\,\cdot\,\ep(p,\sigma)  \,\bmid^2 \spp
\eq
Is there a solution?
Yes, if and only if cuts are not introduced.
In that case the interference terms between different
helicities oscillate over the phase space and drop out,
\ie we achieve factorization,
see \Brefs{Uhlemann:2008pm}.
Furthermore,
the MPE should be understood as an ``asymptotic expansion'', see
\Brefs{Nekrasov:2007ta,Tkachov:1999qb},
not as a narrow-width approximation (NWA).
The phase space
decomposition is obtained by using the two parts
in the propagator expansion of \eqn{fact}:
the $\delta\,$-term is what we need to reconstruct POs,
the PV-term gives the remainder,
and POs
are extracted without making any approximation.
It is worth noting that in extracting
pseudo-observables,
analytic continuation
(of on-shell masses into complex poles) is performed only
after integrating over residual variables~\cite{Goria:2011wa}.

The MPE returns Green's functions in well-defined kinematic limits,
\ie residues of the poles after extracting
the parts which are one-particle-reducible.
These residues can then be computed within SMEFT (or any BSM model)
and expressed in terms of Wilson coefficients (or BSM Lagrangian parameters).

We can illustrate the MPE-PO connection
by using a simple but non-trivial example:
the Dalitz decay of the Higgs boson, see \Bref{Passarino:2013nka}.
Consider the process
\bq
\PH(P) \to \PAf(p_1) + \Pf(p_2) + \PGg(p_3) \spc
\label{prct}
\eq
and introduce invariants
$s_{\sPH}= - P^2$,
$s = - \lpar p_1 + p_2 \rpar^2$,
and propagators $\proA{i} = 1/s_i$
and $\proZ{i} = 1/(s_i - s_{\sPZ})$.
With $s_{\sPH} = \muhs - i\,\muh\,\gh$
we denote the $\PH$ complex pole, \etc.
In the limit of $m_{\Pf} \to 0$,
the total amplitude for process \eqn{prct}
is given by the sum of three contributions:
$\PZ\,$-resonant,
$\PA\,$-resonant,
and non-resonant,
\bq
\mcA\lpar \PH \to \PAf\Pf\PGg\rpar =
 \Bigl[ \mrA^{\mu}_{\sPZ}\lpar \cph\,,\,s \rpar\,\proZ{s} +
        \mrA^{\mu}_{\sPA}\lpar \cph\,,\,s \rpar\,\proA{s} \Bigr]\,e_{\mu}\lpar p_3,l\rpar +
 \mrA_{\NR} \spc
\label{dect}
\eq
where $e_{\mu}$ is the photon polarization vector.
The two resonant components are given by
\bq
\mrA^{\mu}_{\sPV}\lpar \cph\,,\,s \rpar =
\mcT_{\sPHAV}\lpar \cph\,,\,s \rpar\,\mrT^{\mu}_{\nu}\lpar q\,,\,p_3 \rpar\,
J^{\nu}_{\PV\,\Pf}\lpar q\,;\,p_1,p_2\rpar \spc
\eq
where $J^{\mu}_{\PV\,\Pf}$
is the $\PV$-fermion ($\Pf$) current,
$\mrT^{\mu\nu}\lpar k_1\,,\,k_2 \rpar =
\spro{k_1}{k_2}\,\delta^{\mu\nu} - k^{\nu}_1\,k^{\mu}_2$,
$q = p_1 + p_2$,
and
$\PV = \PA,\PZ$
.
Having the full amplitude,
we start the MPE according to
\bq
\mcT_{\sPHAZ}\lpar \cph\,,\,s \rpar =
\mcT_{\sPHAZ}\lpar \cph\,,\,\cpz \rpar +
\lpar s - \cpz \rpar\,\mcT^{(1)}_{\sPHAZ}\lpar \cph\,,\,s \rpar
\qquad \mbox{\etc,}
\eq
and obtain the following result:
\bqa
\mcA\lpar \PH \to \PAf\Pf\PGg\rpar &=& \mrT_{\mu\nu}\lpar q\,,\,p_3 \rpar\,\Bigl[
\mcT_{\sPHAZ}\lpar \cph\,,\,\cpz \rpar\,\proZ{s}\,J^{\nu}_{\PZ\,\Pf}\lpar q\,;\,p_1,p_2\rpar
+ \mcT_{\sPHAA}\lpar \cph\,,\,0 \rpar\,\proA{s}\,J^{\nu}_{\PA\,\Pf}\lpar q\,;\,p_1,p_2\rpar
\nl
{}&+& \mcT^{(1)}_{\sPHAZ}\lpar \cph\,,\,s \rpar\,J^{\nu}_{\PZ\,\Pf}\lpar q\,;\,p_1,p_2\rpar
+ \mcT^{(1)}_{\sPHAA}\lpar \cph\,,\,s \rpar\,J^{\nu}_{\PA\,\Pf}\lpar q\,;\,p_1,p_2\rpar
      \Bigr]\,e^{\mu}\lpar p_3,l\rpar +
      \mrA_{\NR}
\eqa
It is easy to verify that
\bq
\sum_{\lambda=0\,,\,\pm 1}\,e_{\mu}\lpar q\,,\,\lambda \rpar\,e^*_{\nu}\lpar q\,,\,\lambda \rpar =
\delta_{\mu\nu} + \frac{1}{s}\,q_{\mu}\,q_{\nu} \spc
\qquad \spro{q}{q} = - s \spp
\eq
Consider now the single-resonant, $\PZ$, part
\bq
\mcA_{\SR\,;\,\sPAZ} = \mcT_{\sPHAZ}\lpar \cph\,,\,\cpz \rpar\,
                       \mrT_{\mu\nu}\lpar q\,,\,p_3 \rpar\,\proZ{s}\,
                       J^{\nu}_{\PZ\,\Pf}\lpar q\,;\,p_1,p_2\rpar\,e^{\mu}(p_3,l)
\eq
and introduce
\bq
\mrE_i(q) = J^{\mu}_{\PZ\,\Pf}\lpar q\,;\,p_1,p_2\rpar\,e^*_{\mu}(q,i) \spc
\qquad
\mrP_{\mu\,,\,i}(q) = \mcT_{\sPHAZ}\lpar \cph\,,\,\cpz \rpar\,
            \mrT_{\mu\nu}\lpar q\,,\,p_3 \rpar\,
            e^{\nu}(q,i) \spp
\eq
Squaring and summing over spins gives
\bq
\sum_{\spin}\,\bmidl \mcA_{\SR\,;\,\sPAZ} \bmidr^2 =
     \mrP_{\mu\,,\,i}(q)\,\Bigl[ \mrP^{\mu}_j(q) \Bigr]^{\dagger}\,
     \mrE_i(q)\,\mrE^{\dagger}_j(q)\,
     \bmidl \proZ{s} \bmidr^2 \spc
\eq
instead of what is expected for a factorized term, namely
\bq
\sum_{\spin}\,\bmidl \mcA^{\fact}_{\SR\,;\,\sPAZ} \bmidr^2 = \frac{1}{3}\,\sum_{ij}\,
     \mrP_{\mu\,,\,i}(q)\,\Bigl[ \mrP^{\mu}_i(q) \Bigr]^{\dagger}\,
     \bmidl \mrE_j(q) \bmidr^2\,
     \bmidl \proZ{s} \bmidr^2 \spp
\label{wie}
\eq
The result in \eqn{wie} is what we need to define the relevant PO, namely
$\Gamma\lpar \PH \to \PZ\PGg\rpar$.
Derivation continues by writing
\bqs
\Gamma_{\SR}\lpar \PH \to \PAf\Pf\PGg \rpar =
   \frac{1}{2\,\mh}\,\frac{1}{(2\,\pi)^5}\,\int d{\mathrm{PS}}_{1 \to 3}\,
\sum_{\spin}\,\Bigl[ \bmidl \mcA^{\fact}_{\SR\,;\,\sPAZ} \bmidr^2 +
\Delta \mcA_{\SR\,;\,\sPAZ} \Bigr] \spc
\eqs
where $
\bmidl \mcA_{\SR\,;\,\sPAZ} \bmidr^2 =
\bmidl \mcA^{\fact}_{\SR\,;\,\sPAZ} \bmidr^2 +
\Delta \mcA_{\SR\,;\,\sPAZ}$.
Let us now turn to the phase-space integral for
the process in \eqn{prct}.
With $P^2= - \mhs$, we introduce the Mandelstam variables:
$s = - (P - p_3)^2 = - (p_1 + p_2)^2$,
$t = - (P - p_1)^2 = - (p_2 + p_3)^2$,
and
$u = - (P - p_2)^2 = - (p_1 + p_3)^2$,
such that $s + t + u= \mhs$.
Then consider the $n$-dimensional integral
\bq
\Phi_n(s\,,\,t) = \int \prod_i\, d^n p_i\,\theta(p^0_i)\,\delta(p^2_i)\,
\delta^n(P - \sum_i\,p_i)\,
\delta((P - p_3)^2 + s)\,\delta((p_2 + p_3)^2 + t) \spc
\eq
that is related to the phase-space integral by
\bq
\int d{\mathrm{PS}}_{1 \to 3} = \int_{-\infty}^{+\infty}\, ds dt \Phi_n(s\,,\,t) \spp
\label{ktrick}
\eq
Using \eqn{ktrick} as well as
$0 \le s \le \mhs$,
$0 \le t \le \mhs$, and
$0 \le s + t \le \mhs$,
we can write
\bqa
\Gamma_{\SR}\lpar \PH \to \PAf\Pf\PGg \rpar &=&
   \frac{1}{2}\,\mh^5\,\frac{1}{(2\,\pi)^5}\,\int_0^1 dx_s\,\int_0^{1 - x_s} dx_t\,
   \Phi_4(x_s\,,\,x_t)\,
   \sum_{\spin}\,\Bigl[ \bmidl \mcA^{\fact}_{\SR\,;\,\sPAZ} \bmidr^2 +
   \Delta \mcA_{\SR\,;\,\sPAZ} \Bigr] \spc
\eqa
where we have introduced scaled variables, $s= x_s\,\mhs$, \etc

It is easily seen that
$\Delta \mcA_{\SR\,;\,\sPAZ}$ vanishes
after integration over $0 \le x \le 1$,
but this is not the case if cuts are introduced.
This result is an explicit example
of a general proof given in \Bref{Uhlemann:2008pm}.
We therefore derive the result in the extrapolated scenario.
To summarize the steps,
we have the following:

\paragraph{The \texorpdfstring{$\PZ$}{Z} single-resonant amplitude}
This is given by
\bqa
{}&{}& \mcT_{\sPHAZ}\lpar \cph\,,\,\cpz \rpar\,
\mrT_{\mu\nu}\lpar q\,,\,p_3 \rpar\,
e^{\nu}(p_3\,,\,l)\,\delta^{\mu}_{\alpha}\,
J^{\alpha}_{\PZ\,\Pf}\lpar p\,;\,q,k\rpar\,\proZ{s}
\quad \to
\nl
{}&{}& \sum_i\,
\mcT_{\sPHAZ}\lpar \cph\,,\,\cpz \rpar\,
\mrT_{\mu\nu}\lpar q\,,\,p_3 \rpar\,
e^{\nu}(p_3\,,\,l)\,e^{\mu}(q\,,\,i)\,\Bigl[ e_{\alpha}(q\,,\,i) \Bigr]^{\dagger}
J^{\alpha}_{\PZ\,\Pf}\lpar p\,;\,q,k\rpar\,\proZ{s} \spp
\eqa

\paragraph{The fully extrapolated scenario}
This allows to replace the (squared) $\mrS\,$-matrix
element with
\bq
\sum_{ij}\,
\bmidl
\mcT_{\sPHAZ}\lpar \cph\,,\,\cpz \rpar\,
\mrT_{\mu\nu}\lpar q\,,\,p_3 \rpar\,
e^{\nu}(p_3\,,\,j)\,e^{\mu}(q\,,\,i) \bmidr^2\,
\bmidl \proZ{s} \bmidr^2\,
\frac{1}{3}\,\sum_l\,
\bmidl e_{\alpha}(q\,,\,l)\,
J^{\alpha}_{\PZ\,\Pf}\lpar p\,;\,q,k\rpar \bmidr^2 \spp
\eq
At this point, if cuts are introduced there is an extra contribution.

\paragraph{The decomposition of the resonant part
}
We obtain
\bq
\Gamma_{\SR}\lpar \PH \to \PAf\Pf\PGg \rpar =
\frac{1}{16}\,\frac{1}{(2\,\pi)^5}\,\frac{\pi^2}{\mhc}\,
\int_0^1 dx_s\,\int_0^{1-x_s} dx_t\,
\sum_{ij}\,\bmidl \mrS_{\PH \to \PZ\PGg} \bmidr^2\,
\frac{1}{3}\,\sum_l\,\bmidl \mrS_{\PZ \to \PAf\Pf} \bmidr^2\,
\bmidl \proZb{s} \bmidr^2 \spc
\eq
where the scaled propagator is $\proZb{s} = 1/(x_s - \cpz/\mhs)$.
The integrand does not depend on $x_t$ and we can use
\bq
\int_0^{1 - x_s} dx_t = 1 - x_s \spc
\quad
\int_0^1\,x^n_s \bmidl \proZb{s} \bmidr^2 =
\frac{\pi}{\muZ \gamZ}\,\delta\lpar x_s - \frac{\muZs}{\mhs} \rpar +
\;\;\mbox{reg. part}
\eq
We also introduce
\bq
\mrF_{\proc}\lpar \cpz\,,\,s \rpar = \sum_{\spin}\,\bmidl \mrS_{\proc} \bmidr^2 \spp
\label{sdep}
\eq
The reason for the dependence with $s$ in \eqn{sdep} is due to kinematical factors.
This is to say that the kinematic is real and no approximation is made.

\paragraph{The PO definition}
At this point the POs may be defined as
\bq
\Gamma_{\PO}\lpar \PH \to \PZ\PGg \rpar =
\frac{1}{16\,\pi}\,\frac{1}{\mh}\,\lpar 1 - \frac{\muZs}{\mhs} \rpar\,
\mrF_{\PH \to \PZ\PGg}\lpar \cpz\,,\,\muZs\rpar \spc
\quad
\Gamma_{\PO}\lpar \PZ \to \PAf\Pf \rpar =
\frac{1}{48\,\pi}\,\frac{1}{\muZ}\,
\mrF_{\PZ \to \PAf\Pf}\lpar \cpz\,,\,\muZs\rpar \spp
\eq

\begin{figure}[t]
  \centering

  \includegraphics[width=0.5\textwidth, trim = 50 320 190 80, clip=true]{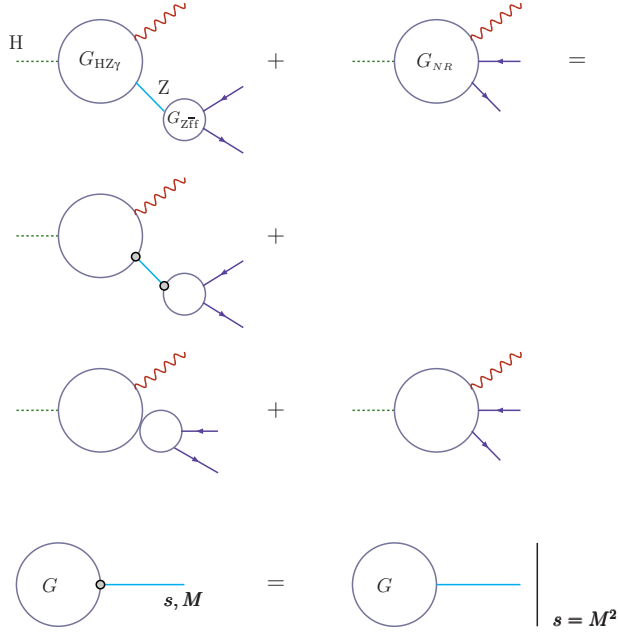}

  \caption{
  Multi-pole-expansion for $\PH \to \PGg \PAf \Pf$.
  $G$ stands for Green's function and $G_{\ssN\ssR}$ denotes
  the non-resonant part of the amplitude.
  The sum of amplitudes in the second (third) row
  is gauge-parameter independent.
  In the last row,
  an amplitude with
  an external line of virtuality $s$ and mass $M$
  is put on-shell.
  \label{fig:HZA}
  }

\end{figure}

\paragraph{The final result}
The final result can be expressed as
\bq
\Gamma_{\SR}\lpar \PH \to \PAf\Pf\PGg \rpar =
\frac{1}{2}\,
\Gamma_{\PO}\lpar \PH \to \PZ\PGg \rpar\,
\frac{1}{\gamZ}\,
\Gamma_{\PO}\lpar \PZ \to \PAf\Pf \rpar + \;\; \mbox{remainder} \spp
\eq
In the narrow-width approximation
the remainder is neglected;
we keep it in our formulation
where the goal is to define POs without making approximations.
Figure~\ref{fig:HZA} illustrates
the MPE of the $\PH \to \PGg \PAf \Pf$ process
as described above.

We can repeat the question: what are POs?
The conclusion is that residues of resonant poles,
$\upkappa\,$-parameters,
and
Wilson coefficients are different layers of POs.
The layer closest to theory refers to
Wilson coefficients
or non-SM parameters in BSM models,
such as
$\alpha, \beta, M_{\mathrm{sb}}$,
\etc in two-Higgs-doublet models.
There is then a layer using kappas,
an intermediate layer defined by residues
that is similar to $g^{\Pe}_{\sPV\,\sPA}$ at LEP,
and a layer closer to experiment
that is similar to $\Gamma(\PZ \to \PAf\Pf)$ at LEP.
However, while
$\Gamma(\PZ \to \PAf\Pf)$ can be defined,
$\Gamma(\PH \to \PZ\PZ)$ cannot
because not all three particles can be on-shell simultaneously;
in other words,
POs are defined by convention,
but they cannot violate kinematics.

One has to be careful
to not confuse the residue
of the two $\PZ$ poles in the
$\PH \to 4\,\Pf$ amplitude
(needed for a question of gauge invariance)
with a partial decay width.
The crucial step is shown in \eqn{fact}:
let $s_{\sPV}$ be the complex pole for a particle $\PV$,
parametrized as $s_{\sPV}= \muVs - i\,\muV \gamV$.
Consider the following integral
\bq
\mrI\lpar a\,,\,b\,,\,s_{\sPV} \rpar = \int_a^b\,d s\frac{1}{\mid s - s_{\sPV} \mid^2} =
     \int_a^b\,d s \frac{1}{(s - \muVs)^2 + \muVs\,\gamVs} \spp
\eq
From \Bref{Actis:2006ra} we obtain
\bqa
{}&{}& \mrI\lpar a\,,\,b\,,\,s_{\sPV} \rpar =
\frac{\pi}{\lpar \mrA\,\lambda \rpar^{1/2}}\,\theta(X)\,\theta(1 - X) +
\mrI^{\reg}\lpar a\,,\,b\,,\,s_{\sPV} \rpar \spc
\nl
{}&{}& \mrI^{\reg}\lpar a\,,\,b\,,\,s_{\sPV} \rpar =
- \frac{1}{2}\,\sum_{l=1,2}\,\bmidl X_l \bmidr^2\,
\int_0^1 dx\,x^{-1/2}\,\lpar \mrA\,X^2_l + \lambda\,x \rpar^{-1} =
- \frac{1}{\mrA}\,\sum_{l=1,2}\,
{}_2\mrF_1\lpar 1\,,\,\frac{1}{2}\,;\,\frac{3}{2}\,;\,- \frac{\lambda}{\mrA X^2_l} \rpar \spc
\nl
{}&{}& {}_2\mrF_1\lpar 1\,,\,\frac{1}{2}\,;\,\frac{3}{2}\,;\, - z^2\rpar =
\frac{1}{z}\,\arctan z \spc
\label{srem}
\eqa
with parameters
\bq
\mrA = b - a \spc
\quad
X = \frac{\muVs - a}{b - a} \spc
\quad
\lambda = \frac{\muVs \gamVs}{b - a} \spc
\quad
X_1 = X_2 = - X \spp
\eq
In general we have
\bq
\mrI_n\lpar a\,,\,b\,,\,s_{\sPV} \rpar = \int_a^b\,d s\frac{s^n}{\mid s - s_{\sPV} \mid^2} =
\frac{\pi}{\lpar \mrA\,\lambda \rpar^{1/2}}\,(\muVs)^n\,\theta(X)\,\theta(1 - X) +
\mrI^{\reg}_n\lpar a\,,\,b\,,\,s_{\sPV} \rpar \spp
\eq
The integral that is needed for the four-body decay,
\bq
I = \int_0^{\mhs} ds_1\,\int_0^{(\mh - \sqrt{s_1})^2} ds_2\,
\lambda^{1/2}\lpar \mhs\,,\,s_1\,,\,s_2 \rpar\,
\bmidl \proZ{s_1}\,\proZ{s_2} \bmidr^2 \spc
\eq
can be worked out along the same lines.
Factorization of phase-space
(the ``opening'' of a line)
requires the identification of
``virtuality with mass''
($s$ with $\muVs$),
which requires $0 \le X \le 1$,
\ie $a \le \muVs \le b$.
Therefore, the natural PO is
$\Gamma(\PH \to \PZ + \PAf \Pf)$.

Some of the POs that were used at LEP
have now been calculated at the two-loop level, see
\Brefs{Hollik:2006ma,Freitas:2014hra,Awramik:2008gi,Awramik:2003ee,Degrassi:2014sxa},
and two-loop renormalization of the full SM has been completed in
\Brefs{Actis:2006ra,Actis:2006rb,Actis:2006rc}.
While the corresponding theoretical uncertainties
are adequate to compare with current precision measurements,
significant improvements will be
necessary to make full use of the precision
foreseen at future facilities, \eg the FCC-ee.
Preliminary studies~\cite{Fccee} seem to indicate
the need for full two-loop exponentiation for QED ISR,
relevant for the measurement itself,
and full three-loop EW radiative corrections,
relevant for the interpretation.
New POs will appear, \eg $\sigma_{\PZ\PH}$
relying on accurate
threshold cross section measurements
sensitive to loop corrections.
\section{What to fit}
In the linear realization of SMEFT,
a subset of $\mrdim= 6$ operators
involves a Higgs doublet $\upPhi$
that contains both the physical Higgs field $\PH$ and
its vacuum expectation value, $\mathrm{v}$.
When using a $\mrdim= 6$ operator
there is a term coming from the replacement of $\upPhi$
with $\mathrm{v}$ (not with $\PH$)
and one gets a shift in $\mrdim= 4$ operators,
\ie kinetic and mass terms.
Normalization of these terms
must always be the canonical one,
\ie the one appearing in the SM Lagrangian.
This means that one has to redefine all fields and parameters,
including the ghost sector,
even before starting the actual calculation of observables.
Furthermore, this set of redefinitions affects the sources
and this must be taken into account when building
$\mathrm{S}\,$-matrix elements out of Green's functions,
for details see \Bref{Ghezzi:2015vva}.
These extra terms are essential
in defining the SMEFT content of all POs.

A question that is often raised
concerns the ``optimal'' parametrization
of the $\mrdim = 6$ basis.
Clearly all bases are equivalent
and there is no obstacle in ``extracting''
(Wilson) coefficients
as defined in a particular basis.
However, certain linear combinations
of Wilson coefficients in one basis
become a single Wilson coefficient in another basis
and a mapping of this type,
that puts coefficients and (pseudo-)observables
in a one-to-one correspondence,
may seem appropriate when considering
LO constraints from EWPD~\cite{Gupta:2014rxa}.
But even at this level one should be careful,
since Wilson coefficients mix under renormalization.
Furthermore, it could be sensible to start with fits at the level
of POs (or kappas),
as usually done in flavor physics,
instead of directly on Wilson coefficients.

Based on these considerations
there are suggestions
on separating weakly- and strongly-constrained
combinations of Wilson coefficients,
possibly disregarding the latter.
However,
this is currently done
in the lowest order implementation
of the experimental constraints
and
there is already strong evidence that
NLO SMEFT provides non-negligible corrections,
which are relevant for per-mille/few percent constraints.
For a given observable $\mcO$
one can compute the deviation
$\mcO_{\SMEFT}/\mcO_{\mySM} - 1$
and the corresponding probability distribution function (pdf)
with the result that the LO pdf differs from the NLO pdf
at the level required by the projected accuracy.

The problem is as follows:
we have a basis where constraints
are not in one-to-one correspondence
with the POs,
which is ideal for the NLO extension (renormalization, mixing
of Wilson coefficients, \etc)
and a basis where LO implementation of constraints is automatic
but (much) less suitable for NLO extension.
The obvious solution is to perform the full NLO SMEFT analysis
in the basis which is well-suited
(therefore reducing the SMEFT theoretical uncertainty),
not only for Higgs physics but also for EWPD,
and only after that one identifies
weakly- and strongly-constrained combinations.

To summarize, mapping of experimental constraints
to Wilson coefficients at LO, and at NLO,
should be corrected for
if an inferred coefficient
is to be used in another process.
Any LO analysis will miss contributions
from the running of Wilson coefficients
(renormalization group)
and from finite ``non-factorizable'' terms
that are not negligible:
as stressed in Sect.~2.4 of \Bref{Berthier:2015gja}
this source of uncertainty
(pertinent to the LO studies)
is not currently included in the fits.
The best way to improve the uncertainty
due to missing higher orders
is to move a step forward
in the perturbative expansion
(both in $g$ and in $\Lambda$).

\section{Effective solutions}

There is now convincing evidence
from the LHC results
that one should
use more theoretical tools;
not only consider more theories (\ie specific models)
but also make use of EFTs.
Both have specific functions,
and both are required~\cite{Hartmann:2001zz}.

The Euler-Heisenberg theory of photon-photon scattering
and the Fermi theory of weak interactions
are prototypical examples of EFTs.
In both theories
only are ``relevant'' fields are considered
and other fields are hidden.
Both theories are valid only up to a scale $\Lambda$,
\eg $E_{\PGg} \muchless m_{\Pe}$,
and unitarity is violated at large scales by Fermi theory.
Both theories are non-renormalizable,
are based on certain symmetries,
and provided stepping stones for scientific advancement.

Also gravity is amenable to an EFT description,
see \Brefs{Donoghue:1995cz,Burgess:2003jk}.
This allows to predict
the effect of quantum physics
on the gravitational interaction
of two heavy masses.
However, such an EFT would only be valid
for ``ordinary'' distances
(where the curvature is small)
and far away from singularities.
Of course, for such a description to be
relevant would require there being no
new physics all the way up to the Planck scale.

%

If one has a full BSM model
it is not necessary to use SMEFT
to describe BSM (Higgs) physics
because
one can always compute anything
from the full BSM model.
However, is very convenient to use SMEFT
because that forces us to concentrate on
universal aspects of SM deviations.
Of course, the SMEFT modifies
the high energy behavior
of any UV completion
and the effective theory
is only a valid description
of the physics at energies below the scale of new physics.
Essentially, interpreting data
via Wilson coefficients
may allow to discern
the UV completion of the SM.
In general, we should strive to devise
a more fundamental description,
since the idea
of an ultimate theory
has a powerful aesthetic pull.
However, do we have that theory?
We have models, mostly ``ad hoc'' models
that cannot be the ``fundamental theory''
and that are sometimes introduced to ``cure''
a specific set of experimental results.
Without entering a detailed discussion,
even if we assume
that a ``fundamental'' theory does exist,
\eg superstrings,
presently we cannot test
its ``resolved'' regime,
\ie phenomena at a very large scale.

This is why SMEFT in the bottom-up approach is so useful:
we do not know what the tower of UV completions is
(or if it exists at all)
but we can formulate the SMEFT
and perform calculations with it
without needing to know what happens
at arbitrarily high scales.

\section{Through the precision straits}

Directly marrying Wilson coefficients
and precision data
to quantify deviations from the SM
is one option,
but not necessarily the most convenient.
Another way of taking the next step is based on POs,
any quantity that is connected to ``data''
by a set of well-defined assumptions.
Properly defining POs requires
care,
since we cannot randomly isolate portions
of an observable and break gauge invariance.
That is why the MPE is useful,
as it provides
(gauge parameter independent) subsets
of the amplitudes.
Of course, residues of resonant poles
can be computed within the SMEFT
in terms of Wilson coefficients,
but that is more akin to
an interpretation step.
Instead, we argue for defining POs
closer to the experimental observations
by defining POs
similar to those at LEP;
the LEP POs have stood the test of time
and still today accurately encode
the information contents
of the data they were derived from.
Introducing such POs and
splitting one observable
into products of POs related to
``sub-processes'' requires
factorization of the full phase-space
and to either
make no cuts (implying an extrapolation)
or compute (and include) correction terms.
%

To conclude,
the journey to the next standard model
may require crossing narrow
straits of precision physics.
If that is what nature has in store for us,
we must equip ourselves with both
a range of concrete BSM models
as well as a general SMEFT.
Both will be
indispensable tools
in navigating an ocean
of future experimental results.
The LEP experience
has proven that
those results can stand the test of time
when expressed in terms of POs.
And as long as POs are well defined
and calculations are performed in
a general and coherent way,
nature can be systematically probed
and
our knowledge of it improved.


\bibliographystyle{elsarticle-num}
\bibliography{wbp2}

\begin{thebibliography}{10}
\expandafter\ifx\csname url\endcsname\relax
  \def\url#1{\texttt{#1}}\fi
\expandafter\ifx\csname urlprefix\endcsname\relax\def\urlprefix{URL }\fi
\expandafter\ifx\csname href\endcsname\relax
  \def\href#1#2{#2} \def\path#1{#1}\fi

\bibitem{Aad:2012tfa}
G.~Aad, et~al., {Observation of a new particle in the search for the Standard
  Model Higgs boson with the ATLAS detector at the LHC}, Phys. Lett. B 716
  (2012) 1--29.
\newblock \href {http://arxiv.org/abs/1207.7214} {\path{arXiv:1207.7214}},
  \href {http://dx.doi.org/10.1016/j.physletb.2012.08.020}
  {\path{doi:10.1016/j.physletb.2012.08.020}}.

\bibitem{Chatrchyan:2012ufa}
S.~Chatrchyan, et~al., {Observation of a new boson at a mass of 125 GeV with
  the CMS experiment at the LHC}, Phys. Lett. B 716 (2012) 30--61.
\newblock \href {http://arxiv.org/abs/1207.7235} {\path{arXiv:1207.7235}},
  \href {http://dx.doi.org/10.1016/j.physletb.2012.08.021}
  {\path{doi:10.1016/j.physletb.2012.08.021}}.

\bibitem{Aad:2015zhl}
G.~Aad, et~al., {Combined Measurement of the Higgs Boson Mass in $pp$
  Collisions at $\sqrt{s}=7$ and 8 TeV with the ATLAS and CMS Experiments},
  Phys. Rev. Lett. 114 (2015) 191803.
\newblock \href {http://arxiv.org/abs/1503.07589} {\path{arXiv:1503.07589}},
  \href {http://dx.doi.org/10.1103/PhysRevLett.114.191803}
  {\path{doi:10.1103/PhysRevLett.114.191803}}.

\bibitem{Bolognesi:2012mm}
S.~Bolognesi, Y.~Gao, A.~V. Gritsan, K.~Melnikov, M.~Schulze, et~al., {On the
  spin and parity of a single-produced resonance at the LHC}, Phys. Rev. D 86
  (2012) 095031.
\newblock \href {http://arxiv.org/abs/1208.4018} {\path{arXiv:1208.4018}},
  \href {http://dx.doi.org/10.1103/PhysRevD.86.095031}
  {\path{doi:10.1103/PhysRevD.86.095031}}.

\bibitem{Aad:2013xqa}
G.~Aad, et~al., {Evidence for the spin-0 nature of the Higgs boson using ATLAS
  data}, Phys. Lett. B 726 (2013) 120--144.
\newblock \href {http://arxiv.org/abs/1307.1432} {\path{arXiv:1307.1432}},
  \href {http://dx.doi.org/10.1016/j.physletb.2013.08.026}
  {\path{doi:10.1016/j.physletb.2013.08.026}}.

\bibitem{Khachatryan:2014kca}
V.~Khachatryan, et~al., {Constraints on the spin-parity and anomalous HVV
  couplings of the Higgs boson in proton collisions at 7 and 8 TeV}, Phys. Rev.
  D 92~(1) (2015) 012004.
\newblock \href {http://arxiv.org/abs/1411.3441} {\path{arXiv:1411.3441}},
  \href {http://dx.doi.org/10.1103/PhysRevD.92.012004}
  {\path{doi:10.1103/PhysRevD.92.012004}}.

\bibitem{Khachatryan:2014jba}
V.~Khachatryan, et~al., {Precise determination of the mass of the Higgs boson
  and tests of compatibility of its couplings with the standard model
  predictions using proton collisions at 7 and 8 TeV}, Eur. Phys. J. C 75~(5)
  (2015) 212.
\newblock \href {http://arxiv.org/abs/1412.8662} {\path{arXiv:1412.8662}},
  \href {http://dx.doi.org/10.1140/epjc/s10052-015-3351-7}
  {\path{doi:10.1140/epjc/s10052-015-3351-7}}.

\bibitem{Aad:2015gba}
G.~Aad, et~al., {Measurements of the Higgs boson production and decay rates and
  coupling strengths using $pp$ collision data at $\sqrt{s}=7$ and $8$ TeV in
  the ATLAS experiment} (2015).
\newblock \href {http://arxiv.org/abs/1507.04548} {\path{arXiv:1507.04548}}.

\bibitem{SomethingOnParameters}
{ALEPH, CDF, D0, DELPHI, L3, OPAL and SLD Collaborations, LEP Electroweak
  Working Group, Tevatron Electroweak Working Group, SLD Electroweak and Heavy
  Flavour Groups}, {Precision electroweak measurements and constraints on the
  Standard Model} (2010).
\newblock \href {http://arxiv.org/abs/1012.2367} {\path{arXiv:1012.2367}}.

\bibitem{Ellis:2013lra}
J.~Ellis, T.~You, {Updated Global Analysis of Higgs Couplings}, JHEP 06 (2013)
  103.
\newblock \href {http://arxiv.org/abs/1303.3879} {\path{arXiv:1303.3879}},
  \href {http://dx.doi.org/10.1007/JHEP06(2013)103}
  {\path{doi:10.1007/JHEP06(2013)103}}.

\bibitem{Englert:2014uua}
C.~Englert, A.~Freitas, M.~M. M{\"u}hlleitner, T.~Plehn, M.~Rauch, M.~Spira,
  K.~Walz, {Precision Measurements of Higgs Couplings: Implications for New
  Physics Scales}, J. Phys. G 41 (2014) 113001.
\newblock \href {http://arxiv.org/abs/1403.7191} {\path{arXiv:1403.7191}},
  \href {http://dx.doi.org/10.1088/0954-3899/41/11/113001}
  {\path{doi:10.1088/0954-3899/41/11/113001}}.

\bibitem{Cranmer:2013hia}
K.~Cranmer, S.~Kreiss, D.~Lopez-Val, T.~Plehn, {Decoupling Theoretical
  Uncertainties from Measurements of the Higgs Boson}, Phys. Rev. D 91~(5)
  (2015) 054032.
\newblock \href {http://arxiv.org/abs/1401.0080} {\path{arXiv:1401.0080}},
  \href {http://dx.doi.org/10.1103/PhysRevD.91.054032}
  {\path{doi:10.1103/PhysRevD.91.054032}}.

\bibitem{Asner:2013psa}
D.~M. Asner, et~al., {ILC Higgs White Paper}, in: {Community Summer Study 2013:
  Snowmass on the Mississippi (CSS2013) Minneapolis, MN, USA, July 29-August 6,
  2013}.
\newblock \href {http://arxiv.org/abs/1310.0763} {\path{arXiv:1310.0763}}.

\bibitem{LHCHiggsCrossSectionWorkingGroup:2012nn}
A.~David, et~al., {LHC HXSWG interim recommendations to explore the coupling
  structure of a Higgs-like particle} (2012).
\newblock \href {http://arxiv.org/abs/1209.0040} {\path{arXiv:1209.0040}}.

\bibitem{Heinemeyer:2013tqa}
S.~Heinemeyer, et~al., {Handbook of LHC Higgs Cross Sections: 3. Higgs
  Properties} (2013).
\newblock \href {http://arxiv.org/abs/1307.1347} {\path{arXiv:1307.1347}},
  \href {http://dx.doi.org/10.5170/CERN-2013-004}
  {\path{doi:10.5170/CERN-2013-004}}.

\bibitem{Contino:2013kra}
R.~Contino, M.~Ghezzi, C.~Grojean, M.~Muhlleitner, M.~Spira, {Effective
  Lagrangian for a light Higgs-like scalar}, JHEP 07 (2013) 035.
\newblock \href {http://arxiv.org/abs/1303.3876} {\path{arXiv:1303.3876}},
  \href {http://dx.doi.org/10.1007/JHEP07(2013)035}
  {\path{doi:10.1007/JHEP07(2013)035}}.

\bibitem{Azatov:2014jga}
A.~Azatov, C.~Grojean, A.~Paul, E.~Salvioni, {Taming the off-shell Higgs
  boson}, Zh. Eksp. Teor. Fiz. 147 (2015) 410--425, [Erratum:
  \DOI{10.7868/S0044451015030039}].
\newblock \href {http://arxiv.org/abs/1406.6338} {\path{arXiv:1406.6338}},
  \href {http://dx.doi.org/10.1134/S1063776115030140}
  {\path{doi:10.1134/S1063776115030140}}.

\bibitem{Contino:2014aaa}
R.~Contino, M.~Ghezzi, C.~Grojean, M.~Muhlleitner, M.~Spira, {eHDECAY: an
  Implementation of the Higgs Effective Lagrangian into HDECAY}, Comput. Phys.
  Commun. 185 (2014) 3412--3423.
\newblock \href {http://arxiv.org/abs/1403.3381} {\path{arXiv:1403.3381}},
  \href {http://dx.doi.org/10.1016/j.cpc.2014.06.028}
  {\path{doi:10.1016/j.cpc.2014.06.028}}.

\bibitem{Berthier:2015oma}
L.~Berthier, M.~Trott, {Towards consistent Electroweak Precision Data
  constraints in the SMEFT}, JHEP 05 (2015) 024.
\newblock \href {http://arxiv.org/abs/1502.02570} {\path{arXiv:1502.02570}},
  \href {http://dx.doi.org/10.1007/JHEP05(2015)024}
  {\path{doi:10.1007/JHEP05(2015)024}}.

\bibitem{Trott:2014dma}
M.~Trott, {On the consistent use of Constructed Observables}, JHEP 02 (2015)
  046.
\newblock \href {http://arxiv.org/abs/1409.7605} {\path{arXiv:1409.7605}},
  \href {http://dx.doi.org/10.1007/JHEP02(2015)046}
  {\path{doi:10.1007/JHEP02(2015)046}}.

\bibitem{Alonso:2013hga}
R.~Alonso, E.~E. Jenkins, A.~V. Manohar, M.~Trott, {Renormalization Group
  Evolution of the Standard Model Dimension Six Operators III: Gauge Coupling
  Dependence and Phenomenology}, JHEP 04 (2014) 159.
\newblock \href {http://arxiv.org/abs/1312.2014} {\path{arXiv:1312.2014}},
  \href {http://dx.doi.org/10.1007/JHEP04(2014)159}
  {\path{doi:10.1007/JHEP04(2014)159}}.

\bibitem{Jenkins:2013wua}
E.~E. Jenkins, A.~V. Manohar, M.~Trott, {Renormalization Group Evolution of the
  Standard Model Dimension Six Operators II: Yukawa Dependence}, JHEP 01 (2014)
  035.
\newblock \href {http://arxiv.org/abs/1310.4838} {\path{arXiv:1310.4838}},
  \href {http://dx.doi.org/10.1007/JHEP01(2014)035}
  {\path{doi:10.1007/JHEP01(2014)035}}.

\bibitem{Jenkins:2013sda}
E.~E. Jenkins, A.~V. Manohar, M.~Trott, {Naive Dimensional Analysis Counting of
  Gauge Theory Amplitudes and Anomalous Dimensions}, Phys. Lett. B 726 (2013)
  697--702.
\newblock \href {http://arxiv.org/abs/1309.0819} {\path{arXiv:1309.0819}},
  \href {http://dx.doi.org/10.1016/j.physletb.2013.09.020}
  {\path{doi:10.1016/j.physletb.2013.09.020}}.

\bibitem{Jenkins:2013zja}
E.~E. Jenkins, A.~V. Manohar, M.~Trott, {Renormalization Group Evolution of the
  Standard Model Dimension Six Operators I: Formalism and lambda Dependence},
  JHEP 10 (2013) 087.
\newblock \href {http://arxiv.org/abs/1308.2627} {\path{arXiv:1308.2627}},
  \href {http://dx.doi.org/10.1007/JHEP10(2013)087}
  {\path{doi:10.1007/JHEP10(2013)087}}.

\bibitem{Jenkins:2013fya}
E.~E. Jenkins, A.~V. Manohar, M.~Trott, {On Gauge Invariance and Minimal
  Coupling}, JHEP 09 (2013) 063.
\newblock \href {http://arxiv.org/abs/1305.0017} {\path{arXiv:1305.0017}},
  \href {http://dx.doi.org/10.1007/JHEP09(2013)063}
  {\path{doi:10.1007/JHEP09(2013)063}}.

\bibitem{Artoisenet:2013puc}
P.~Artoisenet, P.~de~Aquino, F.~Demartin, R.~Frederix, S.~Frixione, et~al., {A
  framework for Higgs characterisation}, JHEP 11 (2013) 043.
\newblock \href {http://arxiv.org/abs/1306.6464} {\path{arXiv:1306.6464}},
  \href {http://dx.doi.org/10.1007/JHEP11(2013)043}
  {\path{doi:10.1007/JHEP11(2013)043}}.

\bibitem{Alloul:2013naa}
A.~Alloul, B.~Fuks, V.~Sanz, {Phenomenology of the Higgs Effective Lagrangian
  via FEYNRULES}, JHEP 04 (2014) 110.
\newblock \href {http://arxiv.org/abs/1310.5150} {\path{arXiv:1310.5150}},
  \href {http://dx.doi.org/10.1007/JHEP04(2014)110}
  {\path{doi:10.1007/JHEP04(2014)110}}.

\bibitem{Ellis:2014dva}
J.~Ellis, V.~Sanz, T.~You, {Complete Higgs Sector Constraints on Dimension-6
  Operators}, JHEP 07 (2014) 036.
\newblock \href {http://arxiv.org/abs/1404.3667} {\path{arXiv:1404.3667}},
  \href {http://dx.doi.org/10.1007/JHEP07(2014)036}
  {\path{doi:10.1007/JHEP07(2014)036}}.

\bibitem{Falkowski:2014tna}
A.~Falkowski, F.~Riva, {Model-independent precision constraints on dimension-6
  operators}, JHEP 02 (2015) 039.
\newblock \href {http://arxiv.org/abs/1411.0669} {\path{arXiv:1411.0669}},
  \href {http://dx.doi.org/10.1007/JHEP02(2015)039}
  {\path{doi:10.1007/JHEP02(2015)039}}.

\bibitem{Low:2009di}
I.~Low, R.~Rattazzi, A.~Vichi, {Theoretical Constraints on the Higgs Effective
  Couplings}, JHEP 04 (2010) 126.
\newblock \href {http://arxiv.org/abs/0907.5413} {\path{arXiv:0907.5413}},
  \href {http://dx.doi.org/10.1007/JHEP04(2010)126}
  {\path{doi:10.1007/JHEP04(2010)126}}.

\bibitem{Degrande:2012wf}
C.~Degrande, N.~Greiner, W.~Kilian, O.~Mattelaer, H.~Mebane, et~al., {Effective
  Field Theory: A Modern Approach to Anomalous Couplings}, Annals Phys. 335
  (2013) 21--32.
\newblock \href {http://arxiv.org/abs/1205.4231} {\path{arXiv:1205.4231}},
  \href {http://dx.doi.org/10.1016/j.aop.2013.04.016}
  {\path{doi:10.1016/j.aop.2013.04.016}}.

\bibitem{Chen:2013kfa}
C.-Y. Chen, S.~Dawson, C.~Zhang, {Electroweak Effective Operators and Higgs
  Physics}, Phys. Rev. D 89 (2014) 015016.
\newblock \href {http://arxiv.org/abs/1311.3107} {\path{arXiv:1311.3107}},
  \href {http://dx.doi.org/10.1103/PhysRevD.89.015016}
  {\path{doi:10.1103/PhysRevD.89.015016}}.

\bibitem{Grober:2015cwa}
R.~Grober, M.~Muhlleitner, M.~Spira, J.~Streicher, {NLO QCD Corrections to
  Higgs Pair Production including Dimension-6 Operators}, JHEP 09 (2015) 092.
\newblock \href {http://arxiv.org/abs/1504.06577} {\path{arXiv:1504.06577}},
  \href {http://dx.doi.org/10.1007/JHEP09(2015)092}
  {\path{doi:10.1007/JHEP09(2015)092}}.

\bibitem{Englert:2015bwa}
C.~Englert, M.~McCullough, M.~Spannowsky, {Combining LEP and LHC to bound the
  Higgs Width} (2015).
\newblock \href {http://arxiv.org/abs/1504.02458} {\path{arXiv:1504.02458}}.

\bibitem{Englert:2015zra}
C.~Englert, I.~Low, M.~Spannowsky, {On-shell interference effects in Higgs
  boson final states}, Phys. Rev. D 91~(7) (2015) 074029.
\newblock \href {http://arxiv.org/abs/1502.04678} {\path{arXiv:1502.04678}},
  \href {http://dx.doi.org/10.1103/PhysRevD.91.074029}
  {\path{doi:10.1103/PhysRevD.91.074029}}.

\bibitem{Biekoetter:2014jwa}
A.~Biekotter, A.~Knochel, M.~Kraemer, D.~Liu, F.~Riva, {Vices and virtues of
  Higgs effective field theories at large energy}, Phys. Rev. D 91 (2015)
  055029.
\newblock \href {http://arxiv.org/abs/1406.7320} {\path{arXiv:1406.7320}},
  \href {http://dx.doi.org/10.1103/PhysRevD.91.055029}
  {\path{doi:10.1103/PhysRevD.91.055029}}.

\bibitem{Gupta:2014rxa}
R.~S. Gupta, A.~Pomarol, F.~Riva, {BSM Primary Effects}, Phys. Rev. D 91~(3)
  (2015) 035001.
\newblock \href {http://arxiv.org/abs/1405.0181} {\path{arXiv:1405.0181}},
  \href {http://dx.doi.org/10.1103/PhysRevD.91.035001}
  {\path{doi:10.1103/PhysRevD.91.035001}}.

\bibitem{Elias-Miro:2013mua}
J.~Elias-Mir{\'o}, J.~Espinosa, E.~Masso, A.~Pomarol, {Higgs windows to new
  physics through d=6 operators: constraints and one-loop anomalous
  dimensions}, JHEP 11 (2013) 066.
\newblock \href {http://arxiv.org/abs/1308.1879} {\path{arXiv:1308.1879}},
  \href {http://dx.doi.org/10.1007/JHEP11(2013)066}
  {\path{doi:10.1007/JHEP11(2013)066}}.

\bibitem{Elias-Miro:2013gya}
J.~Elias-Mir{\'o}, J.~Espinosa, E.~Masso, A.~Pomarol, {Renormalization of
  dimension-six operators relevant for the Higgs decays $h\rightarrow
  \PGg\PGg,\PGg \PZ$}, JHEP 08 (2013) 033.
\newblock \href {http://arxiv.org/abs/1302.5661} {\path{arXiv:1302.5661}},
  \href {http://dx.doi.org/10.1007/JHEP08(2013)033}
  {\path{doi:10.1007/JHEP08(2013)033}}.

\bibitem{Pomarol:2013zra}
A.~Pomarol, F.~Riva, {Towards the Ultimate SM Fit to Close in on Higgs
  Physics}, JHEP 01 (2014) 151.
\newblock \href {http://arxiv.org/abs/1308.2803} {\path{arXiv:1308.2803}},
  \href {http://dx.doi.org/10.1007/JHEP01(2014)151}
  {\path{doi:10.1007/JHEP01(2014)151}}.

\bibitem{Masso:2014xra}
E.~Masso, {An Effective Guide to Beyond the Standard Model Physics}, JHEP 10
  (2014) 128.
\newblock \href {http://arxiv.org/abs/1406.6376} {\path{arXiv:1406.6376}},
  \href {http://dx.doi.org/10.1007/JHEP10(2014)128}
  {\path{doi:10.1007/JHEP10(2014)128}}.

\bibitem{Henning:2014wua}
B.~Henning, X.~Lu, H.~Murayama, {How to use the Standard Model effective field
  theory} (2014).
\newblock \href {http://arxiv.org/abs/1412.1837} {\path{arXiv:1412.1837}}.

\bibitem{Alonso:2014rga}
R.~Alonso, E.~E. Jenkins, A.~V. Manohar, {Holomorphy without Supersymmetry in
  the Standard Model Effective Field Theory}, Phys. Lett. B 739 (2014) 95--98.
\newblock \href {http://arxiv.org/abs/1409.0868} {\path{arXiv:1409.0868}},
  \href {http://dx.doi.org/10.1016/j.physletb.2014.10.045}
  {\path{doi:10.1016/j.physletb.2014.10.045}}.

\bibitem{Hartmann:2015oia}
C.~Hartmann, M.~Trott, {On one-loop corrections in the standard model effective
  field theory; the $\Gamma(h \rightarrow \gamma \, \gamma)$ case}, JHEP 07
  (2015) 151.
\newblock \href {http://arxiv.org/abs/1505.02646} {\path{arXiv:1505.02646}},
  \href {http://dx.doi.org/10.1007/JHEP07(2015)151}
  {\path{doi:10.1007/JHEP07(2015)151}}.

\bibitem{Passarino:1989ey}
G.~Passarino, M.~J.~G. Veltman, {On the Definition of the Weak Mixing Angle},
  Phys. Lett. B 237 (1990) 537.
\newblock \href {http://dx.doi.org/10.1016/0370-2693(90)91221-V}
  {\path{doi:10.1016/0370-2693(90)91221-V}}.

\bibitem{Veltman:1968ki}
M.~J.~G. Veltman, {Perturbation theory of massive Yang-Mills fields}, Nucl.
  Phys. B 7 (1968) 637--650.
\newblock \href {http://dx.doi.org/10.1016/0550-3213(68)90197-1}
  {\path{doi:10.1016/0550-3213(68)90197-1}}.

\bibitem{Buchalla:2013rka}
G.~Buchalla, O.~Cat{\`a}, C.~Krause, {Complete Electroweak Chiral Lagrangian
  with a Light Higgs at NLO}, Nucl. Phys. B 880 (2014) 552--573.
\newblock \href {http://arxiv.org/abs/1307.5017} {\path{arXiv:1307.5017}},
  \href {http://dx.doi.org/10.1016/j.nuclphysb.2014.01.018}
  {\path{doi:10.1016/j.nuclphysb.2014.01.018}}.

\bibitem{Ghezzi:2015vva}
M.~Ghezzi, R.~Gomez-Ambrosio, G.~Passarino, S.~Uccirati, {NLO Higgs effective
  field theory and kappa-framework}, JHEP 07 (2015) 175.
\newblock \href {http://arxiv.org/abs/1505.03706} {\path{arXiv:1505.03706}},
  \href {http://dx.doi.org/10.1007/JHEP07(2015)175}
  {\path{doi:10.1007/JHEP07(2015)175}}.

\bibitem{Hartmann:2015aia}
C.~Hartmann, M.~Trott, {Higgs decay to two photons at one-loop in the SMEFT}
  (2015).
\newblock \href {http://arxiv.org/abs/1507.03568} {\path{arXiv:1507.03568}}.

\bibitem{Actis:2008ts}
S.~Actis, G.~Passarino, C.~Sturm, S.~Uccirati, {NNLO Computational Techniques:
  The Cases $\PH \to \PGg \PGg$ and $\PH \to \Pg \Pg$}, Nucl. Phys. B 811
  (2009) 182--273.
\newblock \href {http://arxiv.org/abs/0809.3667} {\path{arXiv:0809.3667}},
  \href {http://dx.doi.org/10.1016/j.nuclphysb.2008.11.024}
  {\path{doi:10.1016/j.nuclphysb.2008.11.024}}.

\bibitem{Aad:2015tna}
G.~Aad, et~al., {Constraints on non-Standard Model Higgs boson interactions in
  an effective field theory using differential cross sections measured in the
  $H \rightarrow \gamma\gamma$ decay channel at $\sqrt{s} = 8$ TeV with the
  ATLAS detector} (2015).
\newblock \href {http://arxiv.org/abs/1508.02507} {\path{arXiv:1508.02507}}.

\bibitem{Grzadkowski:2010es}
B.~Grzadkowski, M.~Iskrzy{\'n}ski, M.~Misiak, J.~Rosiek, {Dimension-Six Terms
  in the Standard Model Lagrangian}, JHEP 10 (2010) 085.
\newblock \href {http://arxiv.org/abs/1008.4884} {\path{arXiv:1008.4884}},
  \href {http://dx.doi.org/10.1007/JHEP10(2010)085}
  {\path{doi:10.1007/JHEP10(2010)085}}.

\bibitem{Ghezzi:2014qpa}
M.~Ghezzi, G.~Passarino, S.~Uccirati, {Bounding the Higgs Width Using Effective
  Field Theory}, PoS LL2014 (2014) 072.
\newblock \href {http://arxiv.org/abs/1405.1925} {\path{arXiv:1405.1925}}.

\bibitem{Khachatryan:2015mma}
V.~Khachatryan, et~al., {Limits on the Higgs boson lifetime and width from its
  decay to four charged leptons} (2015).
\newblock \href {http://arxiv.org/abs/1507.06656} {\path{arXiv:1507.06656}}.

\bibitem{Ahn:1988fx}
C.~Ahn, M.~E. Peskin, B.~W. Lynn, S.~B. Selipsky, {Delayed Unitarity
  Cancellation and Heavy Particle Effects in $e^+ e^- \to W^+ W^-$}, Nucl.
  Phys. B 309 (1988) 221.
\newblock \href {http://dx.doi.org/10.1016/0550-3213(88)90081-8}
  {\path{doi:10.1016/0550-3213(88)90081-8}}.

\bibitem{Einhorn:2013kja}
M.~B. Einhorn, J.~Wudka, {The Bases of Effective Field Theories}, Nucl. Phys. B
  876 (2013) 556--574.
\newblock \href {http://arxiv.org/abs/1307.0478} {\path{arXiv:1307.0478}},
  \href {http://dx.doi.org/10.1016/j.nuclphysb.2013.08.023}
  {\path{doi:10.1016/j.nuclphysb.2013.08.023}}.

\bibitem{Bardin:1999ak}
D.~Y. Bardin, G.~Passarino, {{The standard model in the making: Precision study
  of the electroweak interactions}}, {Oxford University Press, International
  series of monographs on physics. 104} (1999).

\bibitem{Domenech:2012ai}
O.~Domenech, A.~Pomarol, J.~Serra, {Probing the SM with Dijets at the LHC},
  Phys. Rev. D 85 (2012) 074030.
\newblock \href {http://arxiv.org/abs/1201.6510} {\path{arXiv:1201.6510}},
  \href {http://dx.doi.org/10.1103/PhysRevD.85.074030}
  {\path{doi:10.1103/PhysRevD.85.074030}}.

\bibitem{Altmannshofer:2013foa}
W.~Altmannshofer, D.~M. Straub, {New physics in $B \to K^*\mu\mu$?}, Eur. Phys.
  J. C 73 (2013) 2646.
\newblock \href {http://arxiv.org/abs/1308.1501} {\path{arXiv:1308.1501}},
  \href {http://dx.doi.org/10.1140/epjc/s10052-013-2646-9}
  {\path{doi:10.1140/epjc/s10052-013-2646-9}}.

\bibitem{Degrande:2010kt}
C.~Degrande, J.-M. Gerard, C.~Grojean, F.~Maltoni, G.~Servant, {Non-resonant
  New Physics in Top Pair Production at Hadron Colliders}, JHEP 03 (2011) 125.
\newblock \href {http://arxiv.org/abs/1010.6304} {\path{arXiv:1010.6304}},
  \href {http://dx.doi.org/10.1007/JHEP03(2011)125}
  {\path{doi:10.1007/JHEP03(2011)125}}.

\bibitem{Degrande:2011rt}
C.~Degrande, J.-M. Gerard, C.~Grojean, F.~Maltoni, G.~Servant, {An Effective
  approach to same sign top pair production at the LHC and the forward-backward
  asymmetry at the Tevatron}, Phys. Lett. B 703 (2011) 306--309.
\newblock \href {http://arxiv.org/abs/1104.1798} {\path{arXiv:1104.1798}},
  \href {http://dx.doi.org/10.1016/j.physletb.2011.08.003}
  {\path{doi:10.1016/j.physletb.2011.08.003}}.

\bibitem{Degrande:2013kka}
C.~Degrande, {A basis of dimension-eight operators for anomalous neutral triple
  gauge boson interactions}, JHEP 02 (2014) 101.
\newblock \href {http://arxiv.org/abs/1308.6323} {\path{arXiv:1308.6323}},
  \href {http://dx.doi.org/10.1007/JHEP02(2014)101}
  {\path{doi:10.1007/JHEP02(2014)101}}.

\bibitem{Dawson:2014ora}
S.~Dawson, I.~M. Lewis, M.~Zeng, {Effective field theory for Higgs boson plus
  jet production}, Phys. Rev. D 90~(9) (2014) 093007.
\newblock \href {http://arxiv.org/abs/1409.6299} {\path{arXiv:1409.6299}},
  \href {http://dx.doi.org/10.1103/PhysRevD.90.093007}
  {\path{doi:10.1103/PhysRevD.90.093007}}.

\bibitem{Dawson:2015gka}
S.~Dawson, I.~M. Lewis, M.~Zeng, {Usefulness of effective field theory for
  boosted Higgs production}, Phys. Rev. D 91 (2015) 074012.
\newblock \href {http://arxiv.org/abs/1501.04103} {\path{arXiv:1501.04103}},
  \href {http://dx.doi.org/10.1103/PhysRevD.91.074012}
  {\path{doi:10.1103/PhysRevD.91.074012}}.

\bibitem{deBlas:2014ula}
J.~de~Blas, M.~Ciuchini, E.~Franco, D.~Ghosh, S.~Mishima, et~al., {Global
  Bayesian Analysis of the Higgs-boson Couplings} (2014).
\newblock \href {http://arxiv.org/abs/1410.4204} {\path{arXiv:1410.4204}}.

\bibitem{Berthier:2015gja}
L.~Berthier, M.~Trott, {Consistent constraints on the Standard Model Effective
  Field Theory} (2015).
\newblock \href {http://arxiv.org/abs/1508.05060} {\path{arXiv:1508.05060}}.

\bibitem{Passarino:2012cb}
G.~Passarino, {NLO Inspired Effective Lagrangians for Higgs Physics}, Nucl.
  Phys. B 868 (2013) 416--458.
\newblock \href {http://arxiv.org/abs/1209.5538} {\path{arXiv:1209.5538}},
  \href {http://dx.doi.org/10.1016/j.nuclphysb.2012.11.018}
  {\path{doi:10.1016/j.nuclphysb.2012.11.018}}.

\bibitem{Gonzalez-Alonso:2015bha}
M.~Gonzalez-Alonso, A.~Greljo, G.~Isidori, D.~Marzocca, {Electroweak bounds on
  Higgs pseudo-observables and $h \to 4 \ell$ decays} (2015).
\newblock \href {http://arxiv.org/abs/1504.04018} {\path{arXiv:1504.04018}}.

\bibitem{Grassi:2001bz}
P.~A. Grassi, B.~A. Kniehl, A.~Sirlin, {Width and partial widths of unstable
  particles in the light of the Nielsen identities}, Phys. Rev. D 65 (2002)
  085001.
\newblock \href {http://arxiv.org/abs/hep-ph/0109228}
  {\path{arXiv:hep-ph/0109228}}, \href
  {http://dx.doi.org/10.1103/PhysRevD.65.085001}
  {\path{doi:10.1103/PhysRevD.65.085001}}.

\bibitem{Uhlemann:2008pm}
C.~Uhlemann, N.~Kauer, {Narrow-width approximation accuracy}, Nucl. Phys. B 814
  (2009) 195--211.
\newblock \href {http://arxiv.org/abs/0807.4112} {\path{arXiv:0807.4112}},
  \href {http://dx.doi.org/10.1016/j.nuclphysb.2009.01.022}
  {\path{doi:10.1016/j.nuclphysb.2009.01.022}}.

\bibitem{Nekrasov:2007ta}
M.~Nekrasov, {Modified perturbation theory for pair production and decay of
  fundamental unstable particles}, Int. J. Mod. Phys. A 24 (2009) 6071--6103.
\newblock \href {http://arxiv.org/abs/0709.3046} {\path{arXiv:0709.3046}},
  \href {http://dx.doi.org/10.1142/S0217751X09047673}
  {\path{doi:10.1142/S0217751X09047673}}.

\bibitem{Tkachov:1999qb}
F.~V. Tkachov, {On the structure of systematic perturbation theory with
  unstable fields}, in: {High energy physics and quantum field theory.
  Proceedings, 14th International Workshop, QFTHEP'99, Moscow, Russia, May
  27-June 2, 1999}, pp. 641--645.
\newblock \href {http://arxiv.org/abs/hep-ph/0001220}
  {\path{arXiv:hep-ph/0001220}}.

\bibitem{Goria:2011wa}
S.~Goria, G.~Passarino, D.~Rosco, {The Higgs Boson Lineshape}, Nucl. Phys. B
  864 (2012) 530--579.
\newblock \href {http://arxiv.org/abs/1112.5517} {\path{arXiv:1112.5517}},
  \href {http://dx.doi.org/10.1016/j.nuclphysb.2012.07.006}
  {\path{doi:10.1016/j.nuclphysb.2012.07.006}}.

\bibitem{Passarino:2013nka}
G.~Passarino, {Higgs Boson Production and Decay: Dalitz Sector}, Phys. Lett. B
  727 (2013) 424--431.
\newblock \href {http://arxiv.org/abs/1308.0422} {\path{arXiv:1308.0422}},
  \href {http://dx.doi.org/10.1016/j.physletb.2013.10.052}
  {\path{doi:10.1016/j.physletb.2013.10.052}}.

\bibitem{Actis:2006ra}
S.~Actis, A.~Ferroglia, M.~Passera, G.~Passarino, {Two-Loop Renormalization in
  the Standard Model. Part I: Prolegomena}, Nucl. Phys. B 777 (2007) 1--34.
\newblock \href {http://arxiv.org/abs/hep-ph/0612122}
  {\path{arXiv:hep-ph/0612122}}, \href
  {http://dx.doi.org/10.1016/j.nuclphysb.2007.04.021}
  {\path{doi:10.1016/j.nuclphysb.2007.04.021}}.

\bibitem{Hollik:2006ma}
W.~Hollik, U.~Meier, S.~Uccirati, {The Effective electroweak mixing angle
  $\sin^2\theta_{\mathrm{eff}}$ with two-loop bosonic contributions}, Nucl.
  Phys. B 765 (2007) 154--165.
\newblock \href {http://arxiv.org/abs/hep-ph/0610312}
  {\path{arXiv:hep-ph/0610312}}, \href
  {http://dx.doi.org/10.1016/j.nuclphysb.2006.12.001}
  {\path{doi:10.1016/j.nuclphysb.2006.12.001}}.

\bibitem{Freitas:2014hra}
A.~Freitas, {Higher-order electroweak corrections to the partial widths and
  branching ratios of the $\PZ$ boson}, JHEP 04 (2014) 070.
\newblock \href {http://arxiv.org/abs/1401.2447} {\path{arXiv:1401.2447}},
  \href {http://dx.doi.org/10.1007/JHEP04(2014)070}
  {\path{doi:10.1007/JHEP04(2014)070}}.

\bibitem{Awramik:2008gi}
M.~Awramik, M.~Czakon, A.~Freitas, B.~A. Kniehl, {Two-loop electroweak
  fermionic corrections to $\sin^2\theta^{\PAQb\PQb}_{\mathrm{eff}}$ }, Nucl.
  Phys. B 813 (2009) 174--187.
\newblock \href {http://arxiv.org/abs/0811.1364} {\path{arXiv:0811.1364}},
  \href {http://dx.doi.org/10.1016/j.nuclphysb.2008.12.031}
  {\path{doi:10.1016/j.nuclphysb.2008.12.031}}.

\bibitem{Awramik:2003ee}
M.~Awramik, M.~Czakon, {Complete two loop electroweak contributions to the muon
  lifetime in the standard model}, Phys. Lett. B 568 (2003) 48--54.
\newblock \href {http://arxiv.org/abs/hep-ph/0305248}
  {\path{arXiv:hep-ph/0305248}}, \href
  {http://dx.doi.org/10.1016/j.physletb.2003.06.007}
  {\path{doi:10.1016/j.physletb.2003.06.007}}.

\bibitem{Degrassi:2014sxa}
G.~Degrassi, P.~Gambino, P.~P. Giardino, {The $m_{\PW}-m_{\PZ}$ interdependence
  in the Standard Model: a new scrutiny}, JHEP 05 (2015) 154.
\newblock \href {http://arxiv.org/abs/1411.7040} {\path{arXiv:1411.7040}},
  \href {http://dx.doi.org/10.1007/JHEP05(2015)154}
  {\path{doi:10.1007/JHEP05(2015)154}}.

\bibitem{Actis:2006rb}
S.~Actis, G.~Passarino, {Two-Loop Renormalization in the Standard Model Part
  II: Renormalization Procedures and Computational Techniques}, Nucl. Phys. B
  777 (2007) 35--99.
\newblock \href {http://arxiv.org/abs/hep-ph/0612123}
  {\path{arXiv:hep-ph/0612123}}, \href
  {http://dx.doi.org/10.1016/j.nuclphysb.2007.03.043}
  {\path{doi:10.1016/j.nuclphysb.2007.03.043}}.

\bibitem{Actis:2006rc}
S.~Actis, G.~Passarino, {Two-Loop Renormalization in the Standard Model Part
  III: Renormalization Equations and their Solutions}, Nucl.Phys. B 777 (2007)
  100--156.
\newblock \href {http://arxiv.org/abs/hep-ph/0612124}
  {\path{arXiv:hep-ph/0612124}}, \href
  {http://dx.doi.org/10.1016/j.nuclphysb.2007.04.027}
  {\path{doi:10.1016/j.nuclphysb.2007.04.027}}.

\bibitem{Fccee}
{{First FCC-ee mini-workshop on Precision Observables and Radiative
  Corrections}}, \url{https://indico.cern.ch/e/387296/timetable/\#all.detailed}
  (2015).

\bibitem{Hartmann:2001zz}
S.~Hartmann, {Effective field theories, reductionism and scientific
  explanation}, Stud. Hist. Philos. Mod. Phys. 32 (2001) 267--304.
\newblock \href {http://dx.doi.org/10.1016/S1355-2198(01)00005-3}
  {\path{doi:10.1016/S1355-2198(01)00005-3}}.

\bibitem{Donoghue:1995cz}
J.~F. Donoghue, {Introduction to the effective field theory description of
  gravity}, in: {Advanced School on Effective Theories, Almunecar, Spain, June
  25-July 1, 1995}.
\newblock \href {http://arxiv.org/abs/gr-qc/9512024}
  {\path{arXiv:gr-qc/9512024}}.

\bibitem{Burgess:2003jk}
C.~P. Burgess, {Quantum gravity in everyday life: General relativity as an
  effective field theory}, Living Rev. Rel. 7 (2004) 5--56.
\newblock \href {http://arxiv.org/abs/gr-qc/0311082}
  {\path{arXiv:gr-qc/0311082}}, \href {http://dx.doi.org/10.12942/lrr-2004-5}
  {\path{doi:10.12942/lrr-2004-5}}.

\end{thebibliography}

\end{document}